\documentclass[journal]{IEEEtran}
\ifCLASSINFOpdf
   \usepackage[pdftex]{graphicx}
   \DeclareGraphicsExtensions{.pdf,.jpeg,.jpg,.png}
\else
\fi
\usepackage{multirow}
\ifCLASSOPTIONcompsoc
  \usepackage[caption=false,font=normalsize,labelfont=sf,textfont=sf]{subfig}
\else
  \usepackage[caption=false,font=footnotesize]{subfig}
\fi
\usepackage{url}
\usepackage{amsmath,amssymb}

\begin{document}
%
\title{Deep Face Representations for Differential Morphing Attack Detection}
%
%
%

\author{{Ulrich Scherhag, Christian Rathgeb, Johannes Merkle and Christoph Busch}\vspace{-0.4cm}
	\thanks{\hspace{-0.4cm} U. Scherhag is with the da/sec - Biometrics and Internet Security Research Group, Hochschule Darmstadt, Germany\newline C. Rathgeb  is with the da/sec - Biometrics and Internet Security Research Group, Hochschule Darmstadt, Germany, and the secunet Security Networks AG, Essen, Germany\newline J. Merkle  is with the secunet Security Networks AG, Essen, Germany\newline C. Busch   is with the da/sec - Biometrics and Internet Security Research Group, Hochschule Darmstadt, Germany }}

%
%

\markboth{Transactions on Information Forensics and Security}%
{Scherhag \MakeLowercase{\textit{et al.}}: Deep Face Representations for Differential Morphing Attack Detection}
%



\maketitle

The following paper is a pre-print. The publication is currently under review for IEEE Transactions on Information Forensics and Security (TIFS).
\vspace{0.2in} 

\begin{abstract}
	The vulnerability of facial recognition systems to face morphing attacks is well known. Many different approaches for morphing attack detection (MAD) have been proposed in the scientific literature. However, the MAD algorithms proposed so far have mostly been trained and tested on datasets whose distributions of image characteristics are either very limited (e.g., only created with a single morphing tool) or rather unrealistic (e.g., no print-scan transformation). As a consequence, these methods easily overfit on certain image types and the results presented cannot be expected to apply to real-world scenarios. For example, the results of the latest NIST FRVT MORPH show that the majority of submitted MAD algorithms lacks robustness and performance when considering unseen and challenging datasets.
	
	In this work, subsets of the FERET and FRGCv2 face databases are used to create a realistic database for training and testing of MAD algorithms, containing a large number of ICAO-compliant bona fide facial images, corresponding unconstrained probe images, and morphed images created with four different face morphing tools. Furthermore, multiple post-processings are applied on the reference images, e.g., print-scan and JPEG2000 compression. On this database, previously proposed differential morphing algorithms are evaluated and compared. In addition, the application of deep face representations for differential MAD algorithms is investigated. It is shown that algorithms based on deep face representations can achieve very high detection performance (less than 3\%~\mbox{D-EER}) and robustness with respect to various post-processings. Finally, the limitations of the developed methods are analyzed.
\end{abstract}


%
\IEEEpeerreviewmaketitle

\section{Introduction}\label{sec:introduction}
\begin{figure}[t]
	\hfill
	\subfloat[Subject 1]{\includegraphics[width=0.32\linewidth]{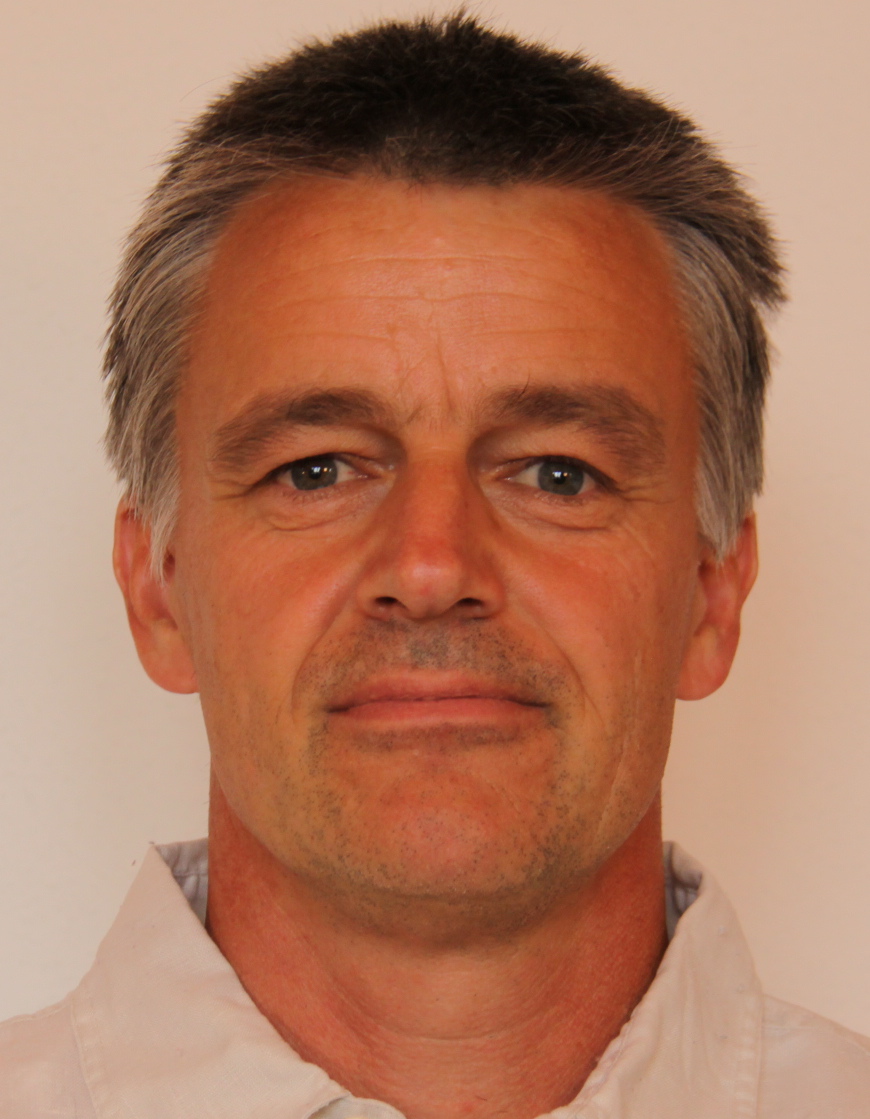}}\hfill
	\subfloat[Morph]{\includegraphics[width=0.32\linewidth]{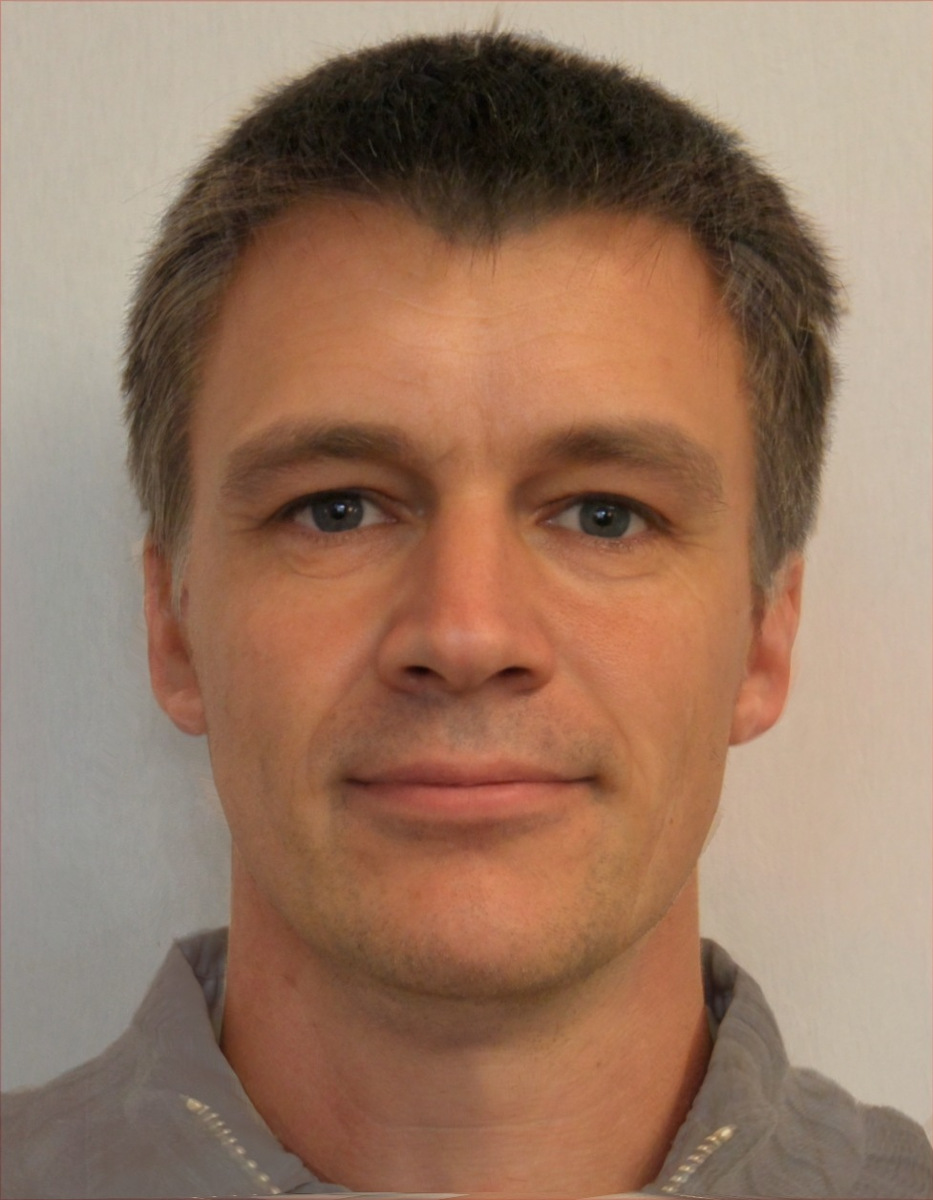}}\hfill	
	\subfloat[Subject 2]{\includegraphics[width=0.32\linewidth]{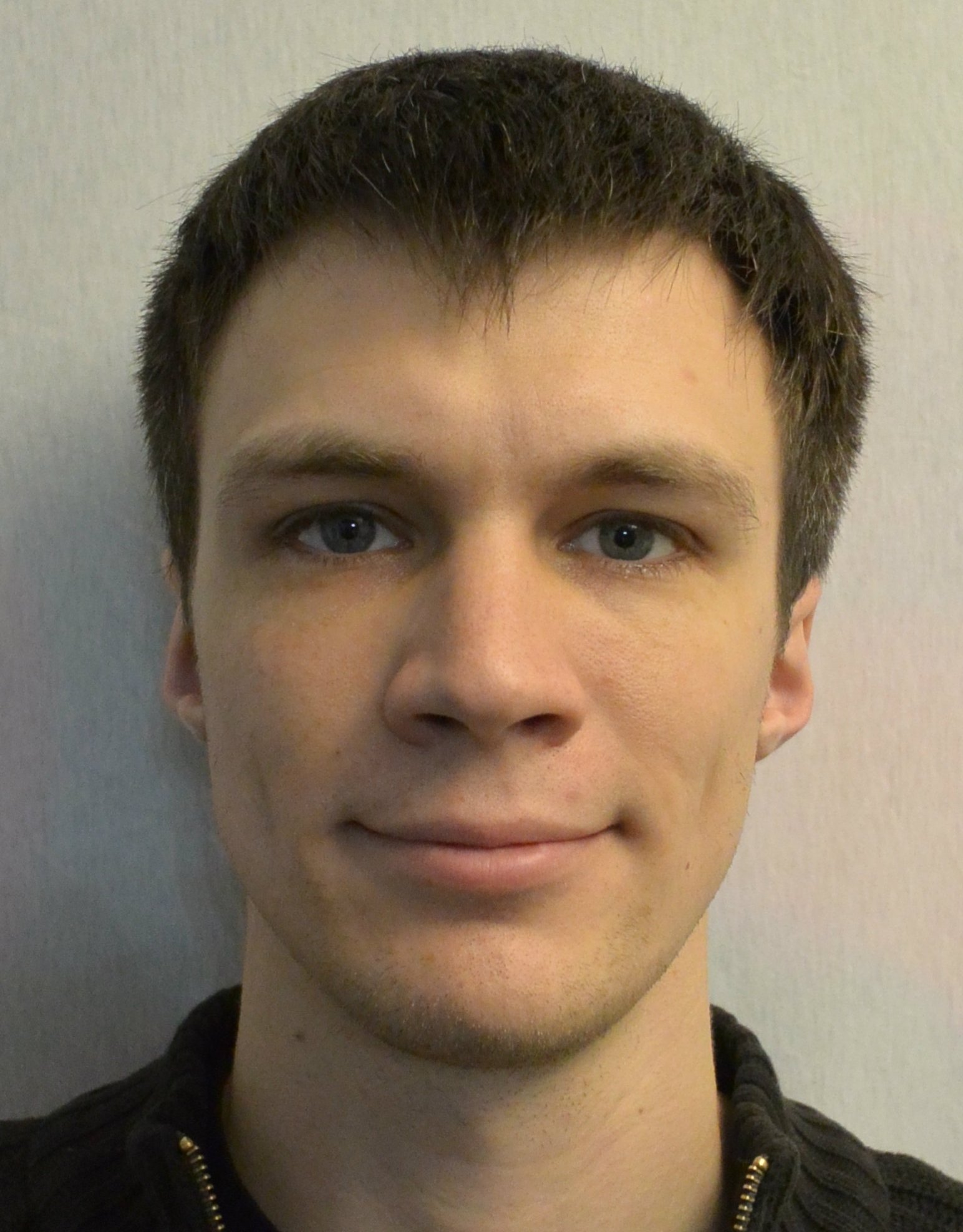}}\hfill
	\caption{Example for a morphed face image (b) of subject 1 (a) and subject 2 (c). The Morph was manually created using FantaMorph.}
	\label{fig:example_morph}
\end{figure}

\IEEEPARstart{I}{mage}  morphing techniques can be used to combine information from two (or more) images into one image. Morphing techniques can also be used to create a morphed facial image from the biometric face images of two individuals, of which the biometric information is similar to that of both individuals. Realistically looking morphed face images can be generated by unskilled users applying readily available tools \cite{Scherhag2019}.  An example of a morphed facial image is shown as part of Figure~\ref{fig:example_morph}.

In many countries, the facial image submitted for an electronic travel document is provided by the applicant either in analogue (i.e., print on paper) or digital form. Therefore, an attacker (e.g., a wanted criminal or a foreigner without authorization to enter a territory) could morph his face image with the face image of a similar looking accomplice who could apply for a passport or another electronic travel document with the morphed image. Since many morphed images are similar enough to deceive human examiners as well as automatic face recognition systems \cite{Ferrara2016, Robertson2018}, the attacker can then use the electronic travel document issued to the accomplice to pass through automatic or manual border controls. The vulnerability of automated face recognition systems against such a Morphing Attack (MA) was initially showcased in \cite{Ferrara2014}. The potential to launch a MA in practice was demonstrated by members of the political activist group \textit{Peng! Kollektiv}, who succeeded without any problem in applying for a passport with a morphed face image\footnote{Peng! Kollektiv, MaskID: \url{https://pen.gg/de/campaign/maskid/}}. Morphing attacks can be applied to other modalities as well. Whether a system is susceptible to this kind of attack can be theoretically examined with the framework presented in \cite{Gomez-Barrero2017} and \cite{Gomez-Barrero2018}.

\setlength{\tabcolsep}{3pt}
\begin{table*}[t]
	\centering
	
	\caption{Overview of most relevant differential MAD algorithms.}\vspace{-0.1cm}
	\label{tab:performance_diff}
	\scriptsize
	\renewcommand*{\arraystretch}{1.2}
	\begin{tabular}{|l|l|l|c|c|c|c|}
		\hline
		\textbf{Reference}	&\textbf{Approach}& \textbf{Category}					& \textbf{Morphing method}	& \textbf{Source face database}	& \textbf{Post-processing}	& \textbf{Remarks}								\\ 
		\hline			
		Scherhag \textit{et al.} \cite{Scherhag2018b}		&	\begin{tabular}{@{}l@{}}Differences in BSIF features \\ with SVM\end{tabular}	&Feature comparison & \begin{tabular}{@{}c@{}}triangulation \\ + blending\end{tabular}	& FRGCv2 \cite{Phillips2005}	& --	& -- \\ \hline
		Scherhag \textit{et al.} \cite{Scherhag2018c}		&	\begin{tabular}{@{}l@{}} Differences in angles of \\ landmark pairs with SVM \end{tabular} & Feature comparison	& \begin{tabular}{@{}c@{}}triangulation \\ + blending\end{tabular}	& ARface \cite{Martinez1998}	& --	& -- \\ \hline
		Damer \textit{et al.} \cite{Damer2018b}			&	\begin{tabular}{@{}l@{}} Directed distances of landmarks\\ with SVM \end{tabular} & Feature comparison & \begin{tabular}{@{}c@{}}triangulation \\ + blending \\ (+ swapping)\end{tabular}	& FERET \cite{Phillips1998}	& --	& -- \\ \hline
		Ferrara \textit{et al.} \cite{Ferrara2018}		&	Demorphing	& Morphing reversion & \begin{tabular}{@{}c@{}}triangulation \\ + blending, GIMP/GAP\end{tabular}	& ARface \cite{Martinez1998}	& --	& --	\\ \hline		
		Ferrara \textit{et al.} \cite{Ferrara2018a}	&	Demorphing & Morphing reversion	& \begin{tabular}{@{}c@{}}triangulation \\ + blending, GIMP/GAP\end{tabular}	& \begin{tabular}{@{}c@{}} ARface \cite{Martinez1998}, \\ CAS-PEAL-R1 \cite{Gao2004} \end{tabular}	& --	& \begin{tabular}{@{}c@{}} CAS-PEAL-R1 contains images\\ with pose variations \end{tabular}	\\ \hline		
		Peng \textit{et al.} \cite{Peng2019}		&	GAN-based Demorphing & Morphing reversion & \begin{tabular}{@{}c@{}}triangulation  \\ + blending \cite{Makrushin2017}\end{tabular}	&	in-house & --	& -- \\ \hline
		
	\end{tabular}
	\vspace{-0.3cm}
\end{table*}

Many approaches for Morphing Attack Detection (MAD) algorithms have already been published in the scientific literature. Most of them can be categorized as {\em single image MAD} algorithms, which examine only the potentially morphed face image, as opposed to {\em differential MAD} methods that compare the image in question with a trusted probe image (e.g., a live capture). While many publications report impressive detection rates, these results are hardly applicable to real-world scenarios. Firstly, the datasets used for evaluation are not realistic. In particular, most publications (\cite{Ferrara2019}, \cite{Ramachandra2017}, \cite{Ramachandra2019} and~\cite{Scherhag2017} being exceptions) do not consider variations in post-processings of the images, e.g., print-scan transformation or severe compression, which can occur in real-world scenarios and may drastically reduce detectable artefacts from the morphing process. In addition, the vast majority of previous publications on differential MAD (with the notable exception of \cite{Ferrara2018a}) use probe images that do not exhibit a realistic intra-subject variance, e.g., facial pose and expression, illumination, the subject's appearance (glasses, beard, hair style, clothing, cosmetics) and aging. Secondly, the datasets used for training and evaluation are not sufficiently distinct but contain images with similar characteristics. In particular, the test sets typically contain only morphed images generated with the same tools as the images in the training set. Consequently, the low error rates reported in these publications may simply reflect an overfitting to the specific and unrealistic properties of the images used. This suspicion is supported by the recent results of the Face Recognition Vendor Test (FRVT) MORPH test~\cite{Ngan2019} conducted by the National Institute of Standards and Technology (NIST), where none of the submitted algorithms achieved satisfactory performance for all test scenarios.  

The main contributions of this work are:

\begin{itemize}
\item A conceptual categorization of published MAD approaches including technical considerations and trade-offs, along with a  more detailed survey of relevant differential MAD schemes. 
\item The creation of a face database containing a large number of ICAO-compliant bona fide and morphed reference images (created using four different morphing algorithms) as well as corresponding unconstrained probe images, which are obtained from different image sources, i.e., face image databases; the use of image post-processings which are likely to be applied in real world, i.e., JPEG2000 compression or print-scan transformations, together with the selection of probe images exhibiting variations in facial pose and expression as well as illumination and focus enable the training and testing of MAD algorithms in a realistic scenario. 
\item The proposal of differential MAD based on deep face representations; by employing state-of-the-art face recognition systems, rich and compact deep facial representations are extracted from pairs of reference and probe images and combined to train machine learning-based classifiers to detect  image alterations induced by morphing algorithms.
\item A comprehensive vulnerability assessment of state-of-the-art face recognition systems against MAs followed by a scenario-based evaluation of the proposed MAD concept applying a commercial and two open-source deep face recognition systems; it is shown that the MAD based on deep face representations outperforms previously proposed differential MAD schemes, which is also confirmed by the latest NIST FRVT MORPH report \cite{Ngan2019}.    
\end{itemize}

The remainder of this work is structured as follows: related works on MAD are summarized in Section~\ref{sec:related}. In Section~\ref{sec:setup} we describe the databases set up for our investigations. Section~\ref{sec:system} describes our approach to develop MAD algorithms based on deep face representations. In Section~\ref{sec:experiments} we evaluate our approach and compare the results with our evaluation of other state-of-the-art MAD algorithms. Finally, conclusions are drawn in Section~\ref{sec:conclusion}.

\section{Related Work}
\label{sec:related}
In recent years, numerous MAD approaches have been proposed. The following subsections give a rough overview of single image and differential MAD algorithms. A more detailed listing and description of the individual algorithms can be found in \cite{Scherhag2019}.

\subsection{Single Image MAD}
Single image MAD approaches can be categorized as \emph{texture descriptors}, e.g., in \cite{Ramachandra2016}, \emph{forensic image analysis}, e.g., in \cite{Kraetzer2017}, and methods based on \emph{deep neural networks}, e.g., in \cite{Seibold2017}. These differ in the artefacts they can potentially detect.

Texture descriptors, e.g., Local Binary Patterns (LBP) \cite{Ojala1996} or Binarized Statistical Image Features (BSIF) \cite{Kannala2012}, attempt to extract discriminative information from images, which can be employed for the purpose of texture classification. The morph process averages the images, which results in smoothed skin textures. Furthermore, ghost artefacts or half-shade effects can occur due to regions that do not overlap exactly (e.g., hair). In particular, in the area of the pupils and nostrils these artefacts appear more frequently, for examples the reader is referred to \cite{Scherhag2017a}. In addition, distorted edges or shifted image areas can occur. These types of artefacts can be easily represented and detected using texture descriptors in multiple ways \cite{Scherhag2018, Scherhag2018a, Scherhag2018b,Ramachandra2019,  Scherhag2017,Ramachandra2016,  Spreeuwers2018, Damer2018a, Ramachandra2017a,  Wandzik2018, Asaad2017, Jassim2018, Agarwal2017}.

Under the assumption that the morphing process leaves specific traces in the image, forensic image analysis techniques can be used to detect them. By averaging the images, the sensor pattern noise is changed. It was shown in \cite{Debiasi2018,Debiasi2018a, Scherhag2019a, Debiasi2019, Zhang2018} that these changes can be used for MAD. Under the assumption that the images are intermediately stored during the morph creation process employing lossy compression algorithms, double compression artefacts can be analyzed \cite{Makrushin2017, Hildebrandt2017}. Furthermore, inconsistencies in the image, e.g., inconsistent illumination \cite{Seibold2018} or color values, might be evaluated.

Deep neural networks can be used to detect morphs in two different ways. Firstly, a new neural network is trained from scratch or an existing neural network is re-trained \cite{Ramachandra2017, Seibold2017, Seibold2018a} for the task of MAD. Deep neural nets can theoretically be trained to detect any artefact. Therefore, it is important that the training data contains a high variance, to avoid overfitting to algorithm and database specific artefacts. Secondly, the feature vectors (embeddings) extracted by existing deep nets can be employed for MAD \cite{Wandzik2018}. Since the neural network was not trained on morphed facial images, it can be assumed that no overfitting to unrealistic morphing artefacts occurs.

\subsection{Differential MAD}
Published differential methods and their properties are listed in Table \ref{tab:performance_diff}. Differential MAD can be divided into two categories. The first category are algorithms that compare feature vectors extracted from extracted from trusted live captures and potential morphs. Single image algorithms can be extended to differential algorithms, e.g., by estimating differences between feature vectors \cite{Scherhag2018b}. The additional information of the trusted live capture might improve the performance and robustness of the detection algorithm. Further, algorithms explicitly utilizing this additional information have been introduced in \cite{Scherhag2018c,Damer2018b}, where the distances between features of facial landmarks are estimated. 

The second category contains algorithms that try to reverse the process of morphing. Here the assumption is that if two subjects are represented in the morphed image and the trusted live capture is subtracted from the possibly morphed reference, in the case of a morphed face image the identity of the second subject becomes more dominant, leading to a lower face recognition scores. If there is no other subject in the image, only the existing one remains. In \cite{Ferrara2018,Ferrara2018a} the so-called demorphing algorithm was proposed, an approach where reversion is done explicitly. In addition to the explicit demorphing approach, demorphing based on a Generative Adversarial Network (GAN) is proposed in \cite{Peng2019}.

\section{Creation of MAD Database}\label{sec:setup}
For the development and evaluation of MAD algorithms, bona fide and morphed reference images are required. In order to resemble passport photos, these images should meet the requirements of the ICAO passport photo quality standards~\cite{ICAO2015}. Multiple tools should be used to generate morphed images to represent a sufficient variance of MAs. In addition, the reference images (bona fide and morphed) should have undergone various realistic post-processings including strong JPG2000 compression and print-scan transformation.  For the investigation of differential MAD, additional probe images are needed. In order to simulate the important use-case of automatic border control, theses probe images should resemble live-captures taken in eGates. Since these recordings are semi-controlled the quality of the captured samples is degraded and does not comply to the ICAO requirements. A much higher variance can be expected, e.g., with respect to pose, facial expression, illumination, and background. Furthermore, as the border control can occur up to ten years after the passport application, reference and probe images can significantly differ with respect to the subject's appearance (glasses, beard, hair style, clothing, cosmetics) and age.

Unfortunately, there is yet no public database available that contains facial images which exhibit all of the mentioned properties. Therefore, we decided to set up a new MAD database based on existing face image databases. In this section, the required steps, including the selection of the images from public face image databases, for the generation of the morphed images, the pairing of reference and probe images, and the application of post-processings to the images, are described in detail.

\subsection{Selection of the Facial Image Database}\label{sec:database}
In a first step, we selected public databases from which we could build our MAD database. The candidate database had to fulfill the following conditions:
\begin{itemize}
	\item It must contain images (meeting the subsequent conditions) of a sufficient amount of different subjects.
	\item The images must have a sufficient resolution. 
    \item For each subject, at least two ICAO-compliant face images must be available (one of which used as input for morph generation and the other as bona fide). 
	\item For each subject, at least one image taken in less constrained conditions (resembling the border control scenario) must be available.
\end{itemize}

Among the available databases, FERET~\cite{Phillips1998} and FRGCv2~\cite{Phillips2005} are suitable. The samples of FERET are all taken in a controlled environment but contain variations in pose and expression. FRGCv2 contains images suitable as passport photos, but also images with scenario variations, e.g., non-uniform illumination, lack of sharpness and uneven background, suitable as probe images.

The FERET and the FRGCv2 face databases were collected over a period of three and two years, respectively. This means the maximum age variation of captured subjects is limited by these periods. That is, selected face images of the used databases exhibit less variations with respect to temporal difference between the capturing of reference and probe images than those expected in a border control scenario. However, to the best of the authors' knowledge, publicly available databases which contain face images captured over a larger period of time do not fulfill the aforementioned requirements. Moreover, we stress that large age variations are only expected to occur for bona fide authentication attempts. For an MA it is more plausible that it will be performed using current face images of the attacker and the accomplice which results in a small temporal difference between capturing of reference and probe images.        

\subsection{Selection of Image Candidates}\label{sec:candidates}
For each of these databases, we select two types of images, representing the reference, i.e., passport photographs, and the unconstrained probe images, i.e., trusted live captures. 

The set of reference images contains all images complying with ICAO requirements~\cite{ICAO2015} except alignment of the face. These images exhibit, among other requirements, uniform illumination, good focus, a neutral face expression with open eyes and no visible teeth, neutral background and no reflections in glasses. For these images, we adjusted the alignment of the face by suitable scaling, rotation and padding/cropping to ensure that the ICAO requirements with respect to the eyes' positions are met. Precisely, facial landmarks are detected applying the Dlib algorithm \cite{King2009} and alignment is performed with respect to the detected eye coordinates where a fixed intra-eye distance of 180 pixels is ensured. 

From the remaining images, those suitable as probe images to resemble the border crossing capture process were selected. The face should be recognizable, but can be only partially illuminated and slightly out of focus. The selected images of the FERET database yield a variance in facial expression and pose (slight rotations). For FRGCv2, the selected images yield a variance in facial expression, background, illumination and sharpness. The probe images are additionally converted to grayscale. This conversion is motivated by the fact that some eGates capture grayscale images. Nonetheless, if face image quality is sufficient, the use of grayscale probe images is expected to have negligible impact on the recognition accuracy of state-of-the-art face recognition systems as well as human examiners \cite{Yip02,Ranjan18a}.

\begin{figure}%
	\centering
	\subfloat[References]{
		\includegraphics[width=0.23\linewidth]{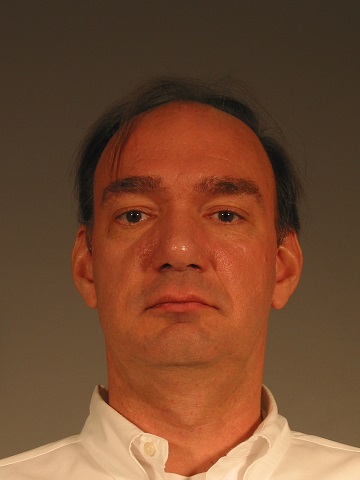}
		\includegraphics[width=0.23\linewidth]{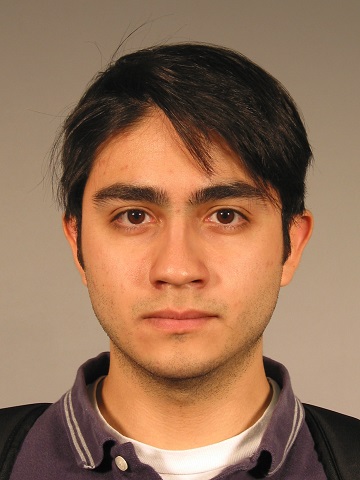}	
		\includegraphics[width=0.23\linewidth]{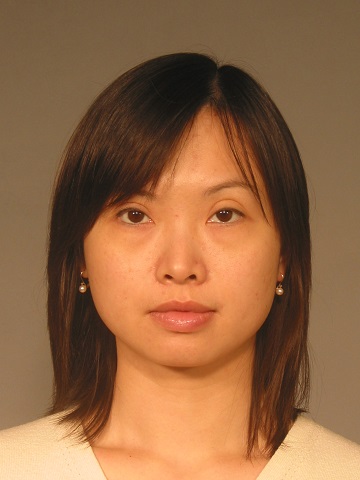}		
		\includegraphics[width=0.23\linewidth]{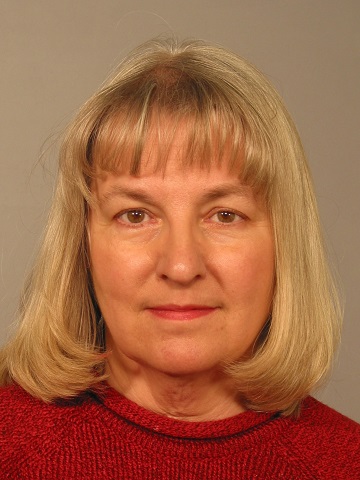}		
	}\\\vspace{-0.2cm}
	\subfloat[Probes]{
		\includegraphics[width=0.23\linewidth]{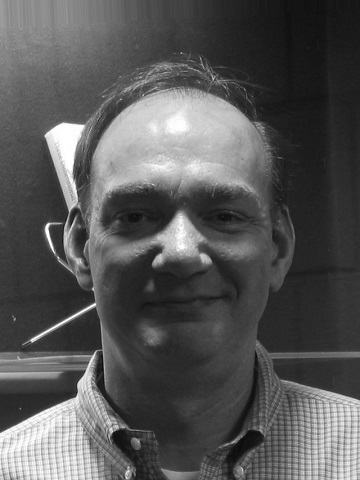}
		\includegraphics[width=0.23\linewidth]{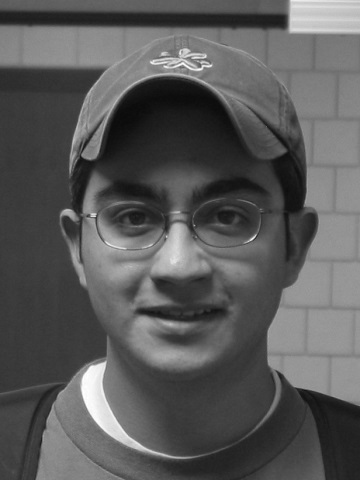}		
		\includegraphics[width=0.23\linewidth]{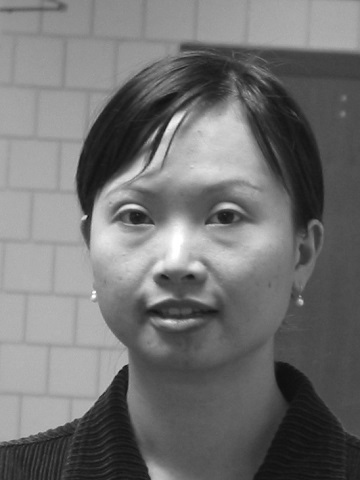}		
		\includegraphics[width=0.23\linewidth]{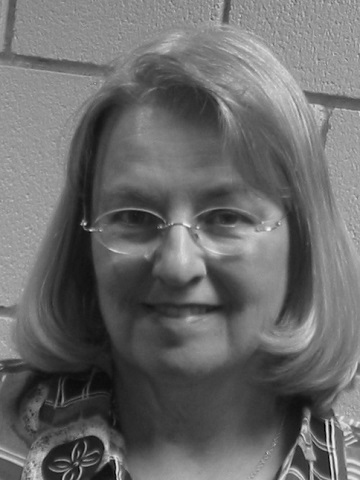}	
	}
	\caption{Examples of reference and gray scale probe images for FRGCv2.}%
	\label{fig:example_probes_frgc}\vspace{-0.2cm}%
\end{figure}

\begin{figure}[t]%
	\centering
	\subfloat[References]{
		\includegraphics[width=0.23\linewidth]{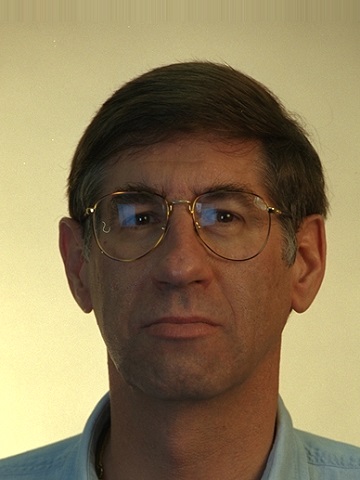}
		\includegraphics[width=0.23\linewidth]{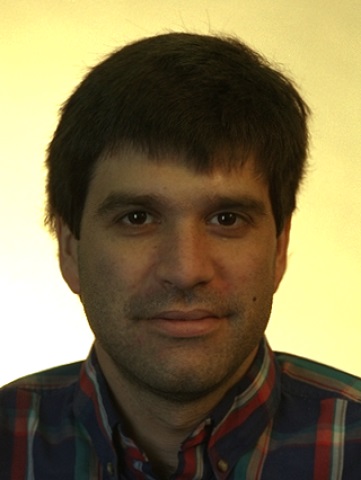}	
		\includegraphics[width=0.23\linewidth]{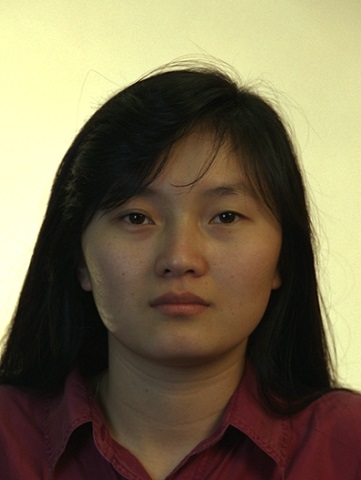}		
		\includegraphics[width=0.23\linewidth]{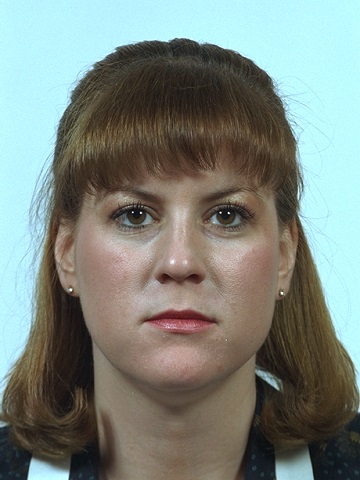}		
	}\\\vspace{-0.2cm}
	\subfloat[Probes]{
		\includegraphics[width=0.23\linewidth]{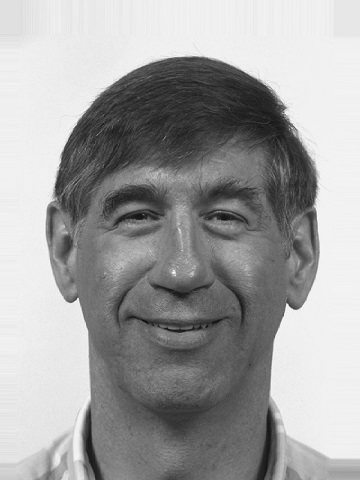}
		\includegraphics[width=0.23\linewidth]{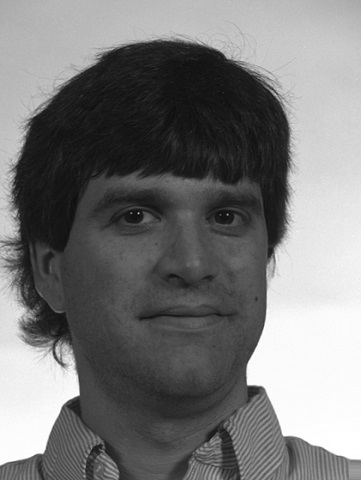}		
		\includegraphics[width=0.23\linewidth]{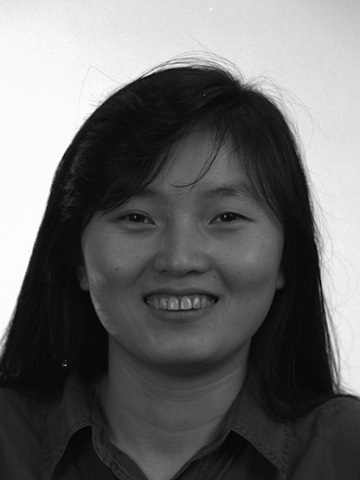}		
		\includegraphics[width=0.23\linewidth]{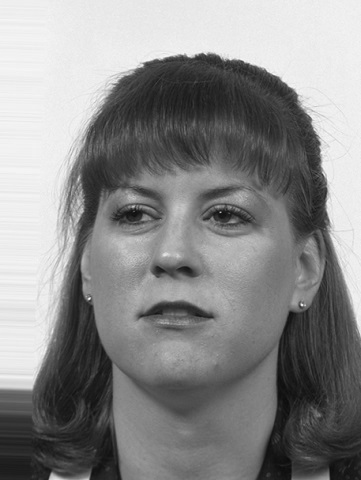}	
	}
	\caption{Examples of reference and gray scale probe images for FERET.}%
	\label{fig:example_probes_feret}\vspace{-0.2cm}%
\end{figure}

Examples of the reference and probe images of both databases are shown in Figure~\ref{fig:example_probes_frgc} and Figure~\ref{fig:example_probes_feret}. Note that the criteria used to select suitable image candidates, which are necessary to resemble a realistic scenario, clearly limit the amount of appropriated images. Eventually, it is important to note that the selection of face images is to a certain degree subjective. In order to facilitate reproducibility of this research the list of selected face images is made available to the research community (upon acceptance). 

\subsection{Selection of Images and Morph Creation}\label{sec:image_pairing}
For each subject, we selected from all available reference images a subset of \emph{bona fide reference images} and a subset of images for generating morphs, i.e., \emph{morph input images}, as well as a subset of \emph{probe images}. Where possible, we ensured that the bona fide reference images were disjoint from the morph input images, thereby avoiding a repetitive use of the same image or same image parts during MAD training which increases the variance of training data. No lossy compression was applied to the images prior to post-processing. In addition, we assured that selected reference images (bona fide or morph input images), if possible, are captured in a different recording session than the probe images, in order to achieve temporal difference between the capturing of reference and probe images as expected in a real border control scenario. For image candidates selected from the FERET database, where possible, we selected one bona fide reference image, one morph input image, and up to two probe images per subject. For image candidates selected from the FRGCv2 database, if possible, we chose two bona fide reference images, two morph input images, and up to five probe images per subject.

For the creation of morphs, the morph input images are lexicographically sorted, and then each input image is morphed with one of the next consecutive input image, for which the following two criteria hold: both depicted subjects are of same sex; only one face image of the image pair shows glasses. The latter criteria should avoid obvious artefacts. Morph input images are only used once, i.e., for the generation of a single morph. It is important to note that after the morph generation the morph input images are discarded. Again, we aim to avoid a repeated use of the same image or same image parts in the training stage of MAD algorithms which should also prevent from overfitting caused by overrepresented image parts. The list of resulting face image pairs used for morph generation is provided to the research community (upon acceptance). The key figures of the resulting database are listed in Table~\ref{tab:database_setup}. 

\begin{table}[t]
	\begin{center}
	\caption{Composition of the MAD database. The number of morph images is multiplied by the number of morphing tools used.}
	\label{tab:database_setup}
	\renewcommand*{\arraystretch}{1.2}
	\begin{tabular}{| l | c | c | c | c | c | c |}
		\hline
		\multirow{2}{*}{\textbf{Database}} & \multirow{2}{*}{\textbf{Subjects}} & \multirow{2}{*}{\textbf{Male}} & \multirow{2}{*}{\textbf{Female}} & \multicolumn{2}{|c|}{\textbf{Reference Images}} & \multirow{2}{*}{\textbf{Probes}} \\
		 &  &  &  & \textbf{Bona Fide} & \textbf{Morphs} &  \\
		\hline
		FERET & 529 & 329 & 200 & 529 & 529$^*$ & 791 \\
		\hline
		FRGCv2 & 533 & 302 & 231 & 984 & 964$^*$ & 1,726 \\
		\hline
	\end{tabular}\newline
	\end{center}
	$^*$per morphing algorithm
\end{table}

\begin{figure*}[!t]
	\begin{minipage}{0.16\linewidth}
		\includegraphics[width=\linewidth]{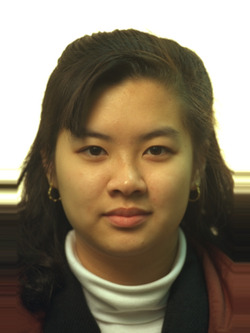}\newline\vspace{-0.3cm}
		\subfloat[Subject 1]{\includegraphics[width=\linewidth]{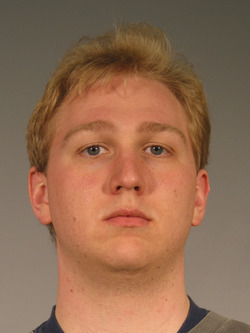} \label{fig:sample_subject1}}
	\end{minipage}
	\begin{minipage}{0.16\linewidth}
		\includegraphics[width=\linewidth]{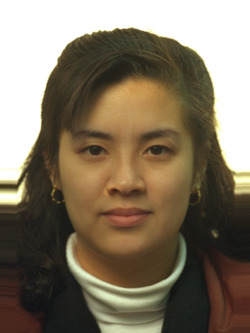}\newline\vspace{-0.3cm}
		\subfloat[FaceFusion]{\includegraphics[width=\linewidth]{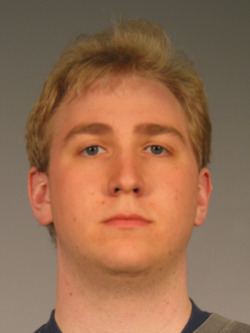} \label{fig:sample_facefusion}}
	\end{minipage}
	\begin{minipage}{0.16\linewidth}
		\includegraphics[width=\linewidth]{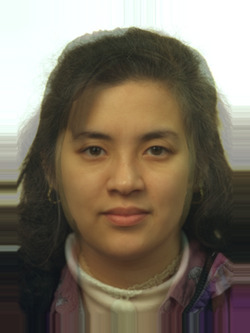}\newline\vspace{-0.3cm}
		\subfloat[FaceMorpher]{\includegraphics[width=\linewidth]{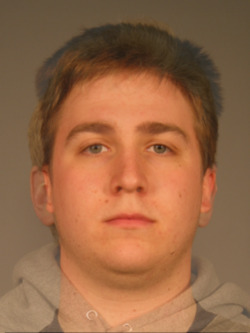} \label{fig:sample_facemorpher}}
	\end{minipage}
	\begin{minipage}{0.16\linewidth}
		\includegraphics[width=\linewidth]{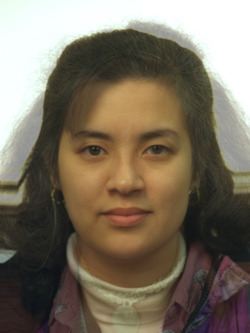}\newline\vspace{-0.3cm}
		\subfloat[OpenCV]{\includegraphics[width=\linewidth]{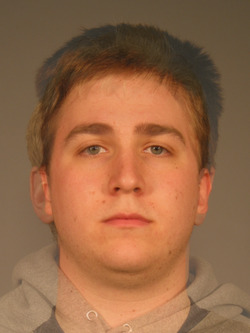} \label{fig:sample_opencv}}
	\end{minipage}
	\begin{minipage}{0.16\linewidth}
		\includegraphics[width=\linewidth]{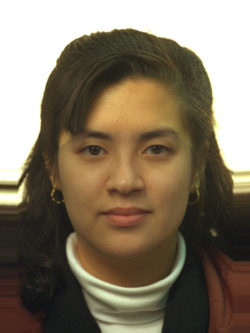}\newline\vspace{-0.3cm}
		\subfloat[UBO-Morpher]{\includegraphics[width=\linewidth]{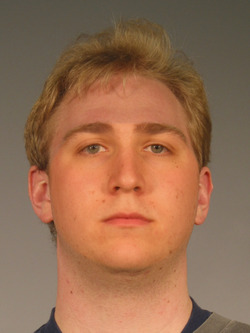} \label{fig:sample_ubo}}
	\end{minipage}
	\begin{minipage}{0.16\linewidth}
		\includegraphics[width=\linewidth]{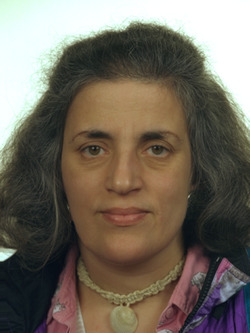}\newline\vspace{-0.3cm}
		\subfloat[Subject 2]{\includegraphics[width=\linewidth]{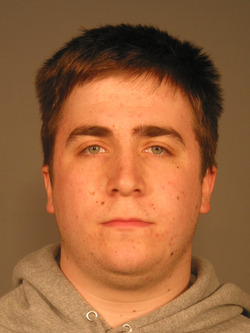} \label{fig:sample_subject2}}
	\end{minipage}	
	\caption{Examples for a morphed face images from all four algorithms for a female pair of face images from the FERET (top row) and the FRGCv2 (bottom row). Equal weights of face images were used to create morphs. Face images depicting subjects with similar demographic attributes, e.g., age and ethnicity, generally result in more plausible face morphs. }
	\label{fig:morph_example}
\end{figure*}
\subsection{Morphing Algorithms}\label{sec:morphing_protocol}
Four different automated morphing tools were used to create morphed face images:

\begin{enumerate}
	\item \textbf{FaceFusion}\footnote{\url{www.wearemoment.com/FaceFusion/}}, a proprietary morphing algorithm. Originally being an iOS app, we deployed an adaptation for Windows which uses the 68 landmarks of Dlib and Delaunay triangles. After the morphing process, certain regions (eyes, nostrils, hair) of the first face image are blended over the morph to hide  artefacts. The corresponding landmarks of upper and lower lips can be reduced as described in~\cite{Makrushin2017} to avoid artefacts at closed mouths. The created morphs have a high quality and low to no visible artefacts. Example morphs are shown in Figure~\ref{fig:sample_facefusion}. 
	\item \textbf{FaceMorpher}\footnote{\url{github.com/alyssaq/face_morpher}}, an open-source implementation using Python. In the version applied for this work, the algorithm uses STASM for landmark detection. Delaunay triangles, which are formed from the landmarks, are wrapped and blended. The area outside the landmarks is averaged. The generated morphs show strong artefacts in particular in the area of neck and hair. Figure~\ref{fig:sample_facemorpher} depicts examples of generated morphs. 
	\item \textbf{OpenCV}, a self-made morphing algorithm derived from ``Face Morph Using OpenCV''\footnote{\url{www.learnopencv.com/face-morph-using-opencv-cpp-python/}}. This algorithm works similar to FaceMorpher. Important differences between the algorithms are that for landmark detection Dlib is used instead of STASM and that additional landmarks are positioned at the edges of the image, which are also used to create the morphs. Thus, in contrast to FaceMorpher, the outer facial area does not consist of an averaged image, but like the rest of the image, of morphed triangles. However, visible artefacts outside the face area can be observed, which is mainly due to missing landmarks. Example morphs can be seen in Figure~\ref{fig:sample_opencv}. 
	\item \textbf{UBO-Morpher}, the morphing tool of University of Bologna, as used, e.g., in \cite{Ferrara2018}. Dlib landmarks were used for this algorithm. The morphs are generated by triangulation, averaging and blending. To avoid the artefacts in the area outside the face, the morphed face is copied to the background of one of the original images. Even if the colors are adjusted at boundaries, visible edges may appear at the transitions. Figure~\ref{fig:sample_ubo} shows examples of resulting morphs.
\end{enumerate}
No manual post-processing is applied in the morph generation process. The use of various morphing tools which produce facial morphs of different quality enable a thorough investigation of MAD capabilities, i.e., impact of the use of high and low quality morphs on training and testing of MAD methods, see Section~\ref{sec:experiments}. The evaluation of MAD algorithms for different quality levels of morphed face images is also performed by NIST in the FRVT MORPH \cite{Ngan2019}. The pairs for the morphing process for each algorithm are selected according the protocol defined in Section~\ref{sec:image_pairing}, resulting in 4$\times$529 morphed face images for FERET and 4$\times$964 morphed face images for FRGCv2. For both databases, the bona fide and morphed face images are normalized to meet the ICAO-requirements for passport images \cite{ICAO2015}. The resulting images are of size 720$\times$960 pixels.

\subsection{Post-Processing}
\label{sec:pre-processing}
\begin{figure}[!t]
	\centering
	\hfil
	\subfloat[NPP]{\includegraphics[width=0.23\linewidth]{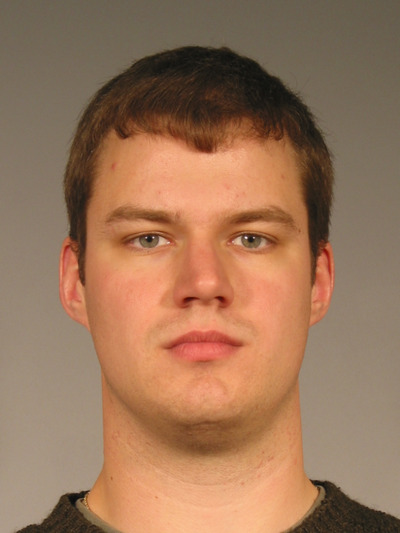} \label{fig:sample_npp}}
	\hfil
	\subfloat[Resized]{\includegraphics[width=0.23\linewidth]{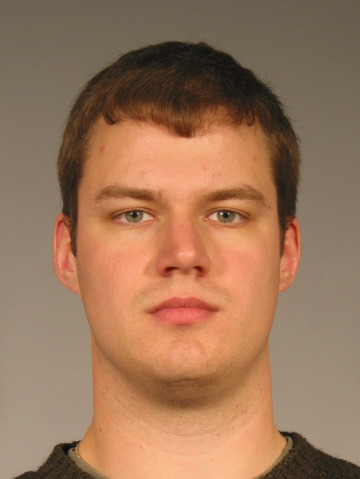} \label{fig:sample_resized}}
	\hfil
	\subfloat[JP2]{\includegraphics[width=0.23\linewidth]{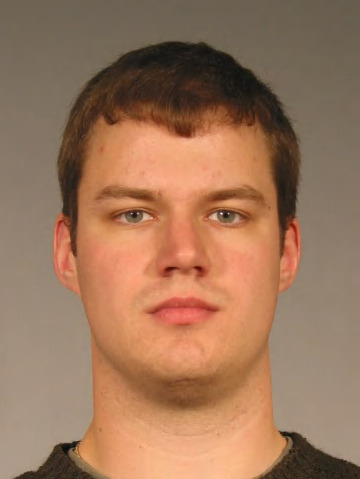} \label{fig:sample_jp}}
	\hfil
	\subfloat[PS-JP2]{\includegraphics[width=0.23\linewidth]{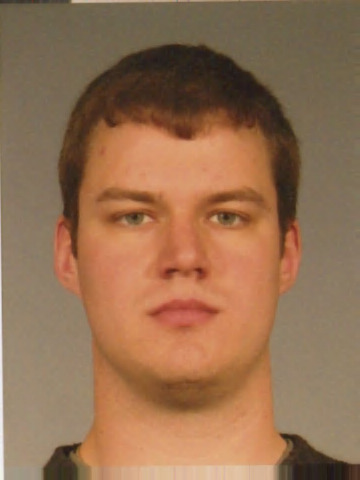} \label{fig:sample_ps300jp}}
	\hfil	
	\caption{Examples of an original image and the three post-processing types.}
	\label{fig:samples_pp}\vspace{-0.4cm}
\end{figure}

Images that have been captured for the use in an identity document, e.g., passport, can go through various processing steps before they are embedded, e.g., in a passport RFID chip. To reflect this variety, the passport images (bona fide and morphed) of the MAD database are post-processed in different manners. An example for the different post-processings is shown in Figure~\ref{fig:samples_pp}.
\begin{quote}
	\textbf{Unprocessed:} The images are not further processed, henceforth referred to as \textit{NPP} (no post-processing). NPP images serves as baseline.\\
	\textbf{Resized:} The resolution of the images is reduced by half, resulting in 360$\times$480 pixels, in the following text referred to as \textit{Resized}. Resized images fulfill the minimum requirement with respect to intra-eye distance defined by ICAO, i.e., 90 pixels. This pre-processing corresponds to the scenario that an image is submitted digitally by the applicant.\\
	\textbf{JPEG2000:} The images are resized by half and then compressed using JPEG2000, a wavelet-based image compression method that is recommended for EU passports \cite{EuropeanCommission2018}. The setting is selected in a way that a target file size of 15KB is achieved. This scenario reflects the post-processing path of passport images if handed over digitally at the application desk, hereafter referred to as \textit{JP2}.\\
	\textbf{Print/Scan -- JPEG2000:} The original images (uncompressed and not resized) are first printed with a high quality laser printer (\textit{Fujifilm Frontier 5700R Minlab} on \textit{Fujicolor Crystal Archive Paper Supreme HD Lustre photo paper}) and then scanned with a premium flatbed scanner (\textit{Epson DS-50000}) with 300 dpi. A dust and scratch filter is then applied in order to reduce image noise. Subsequently, the images are resized to 360$\times$480 pixels, i.e., half of the NPP images, and compressed to 15 KB using JPEG2000.\footnote{Due to the lustre print, the scans exhibit a visible pattern of the paper surface, which is only partly removed by the dust and scratch filter and results in stronger compression artefacts than for scans of glossy prints.} This scenario reflects the post-processing path of passport images if handed over at the application desk as a printed photograph, subsequently referred to as \textit{PS-JP2}.
\end{quote}

\subsection{Validation of Attack Potential}
\label{ssec:vulnerability}
\begin{figure*}[!t]
\centering
	\subfloat[FERET: ArcFace, MAs$_{50}$]{\includegraphics[height=3.25cm]{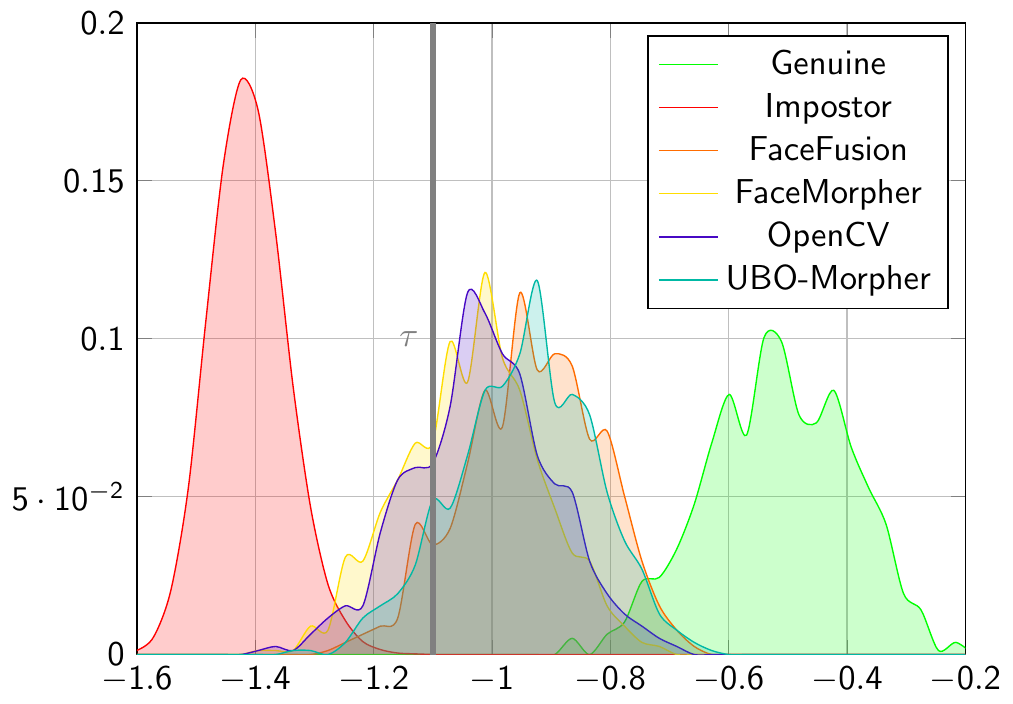}}
	\subfloat[FERET: COTS, MAs$_{50}$]{\includegraphics[height=3.25cm]{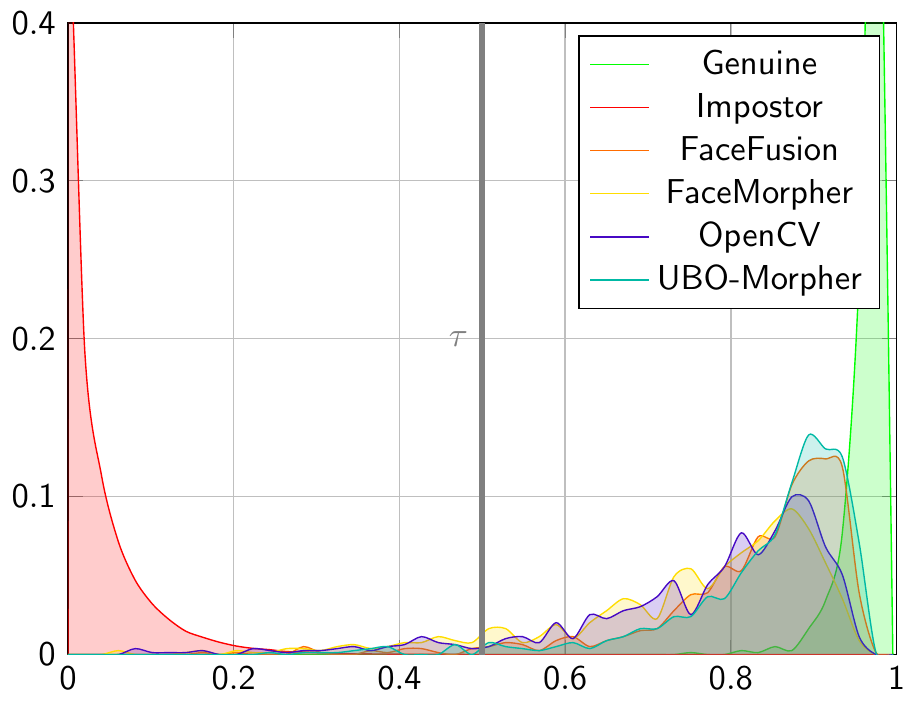}}
	\subfloat[FRGCv2:: ArcFace, MAs$_{50}$]{\includegraphics[height=3.25cm]{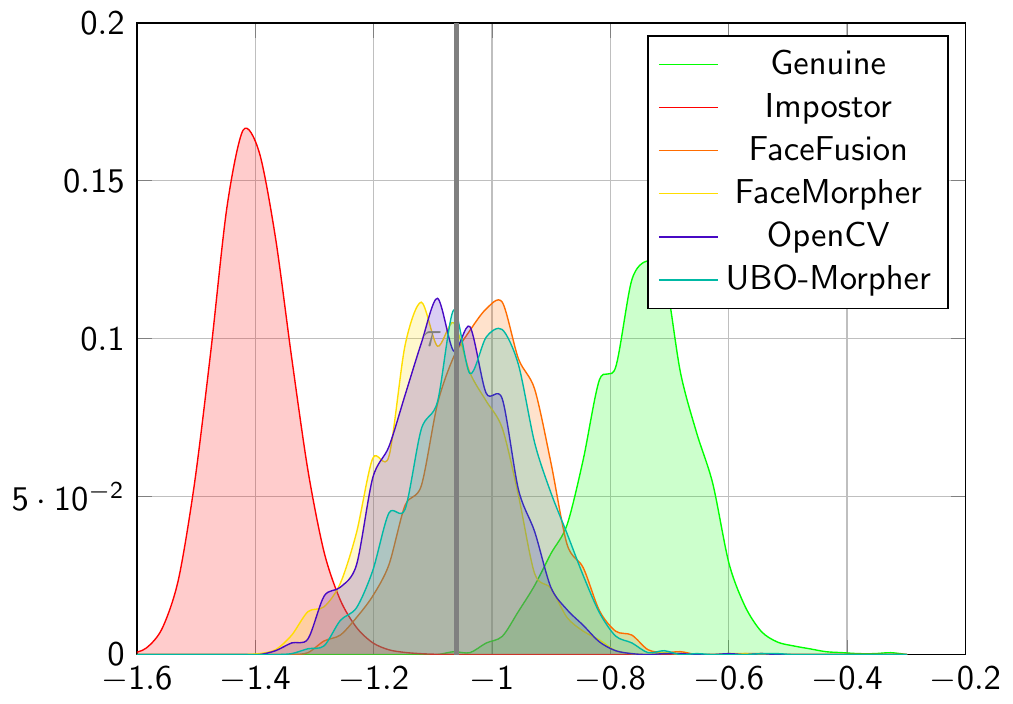}}
	\subfloat[FRGCv2: COTS, MAs$_{50}$]{\includegraphics[height=3.25cm]{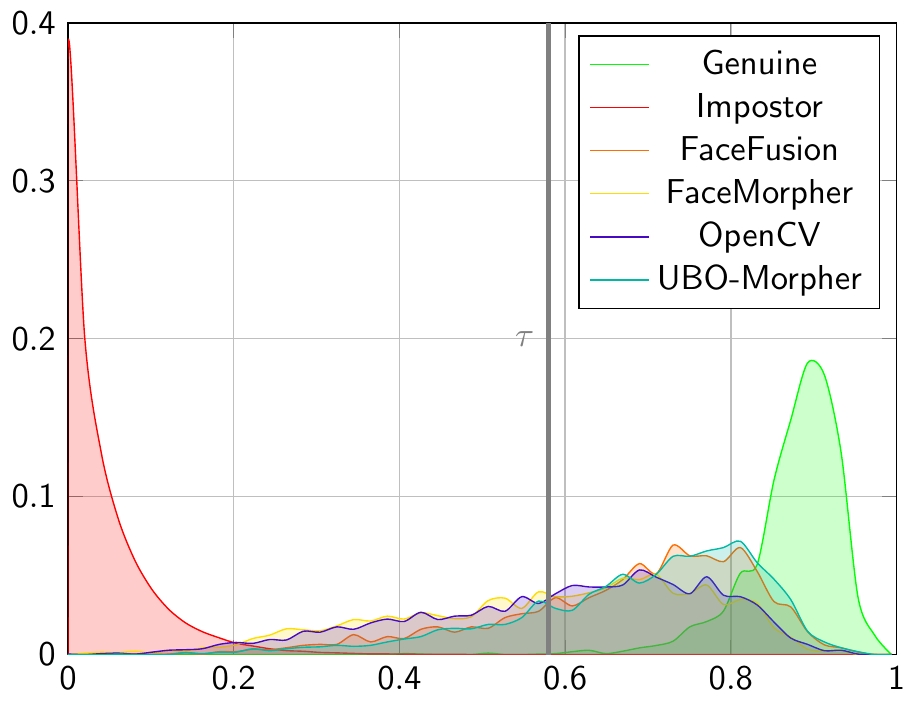}}\\
	\subfloat[FERET: ArcFace, MAs$_{25}$]{\includegraphics[height=3.25cm]{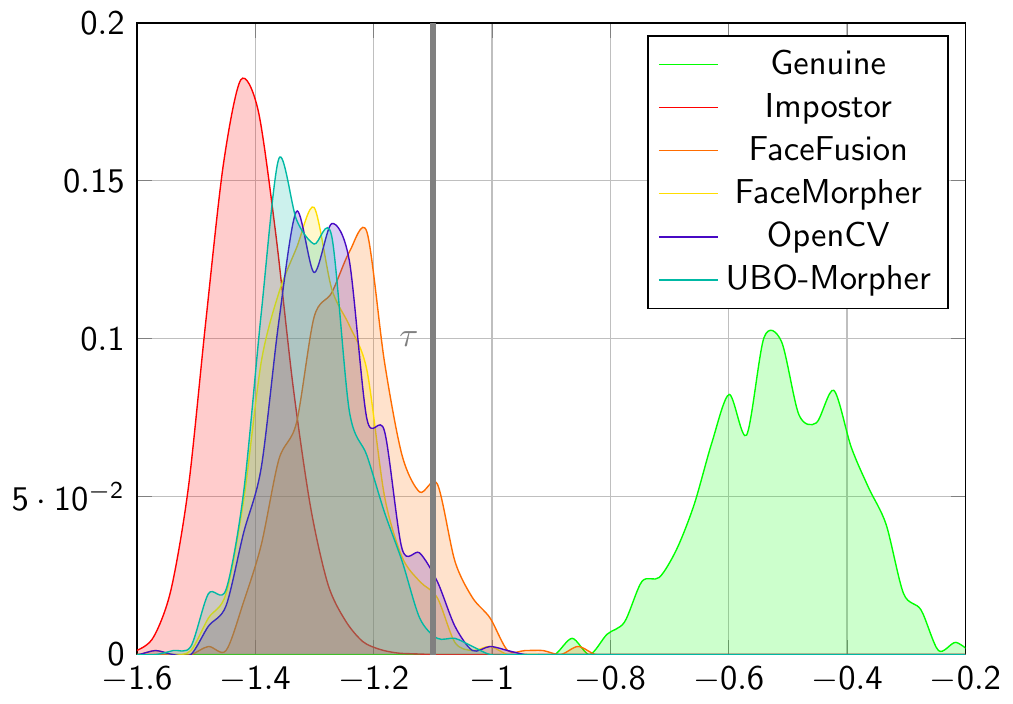}}
	\subfloat[FERET: COTS, MAs$_{25}$]{\includegraphics[height=3.25cm]{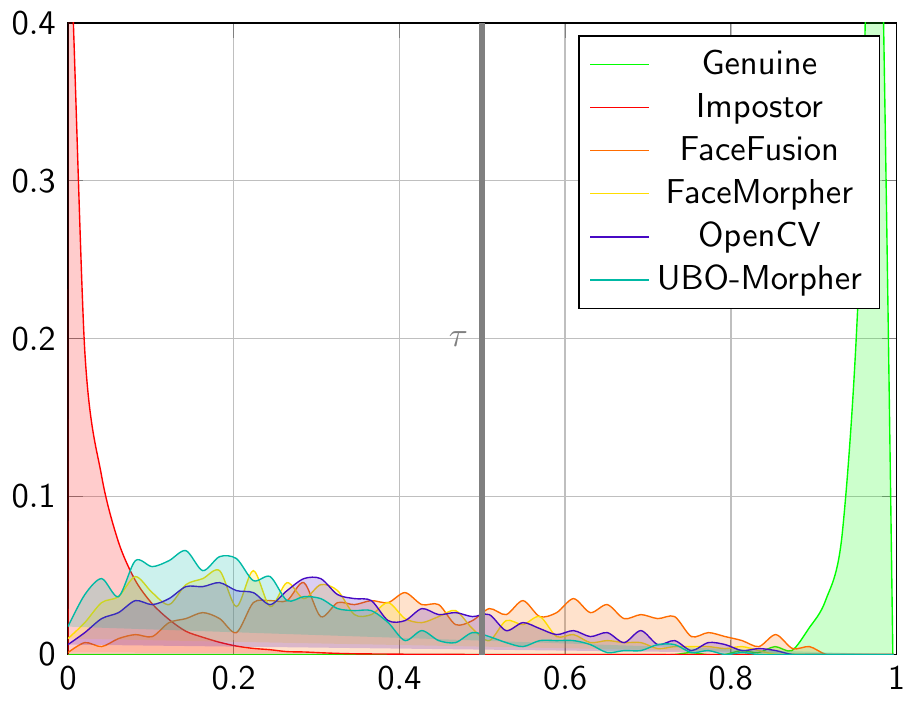}}
	\subfloat[FRGCv2: ArcFace, MAs$_{25}$]{\includegraphics[height=3.25cm]{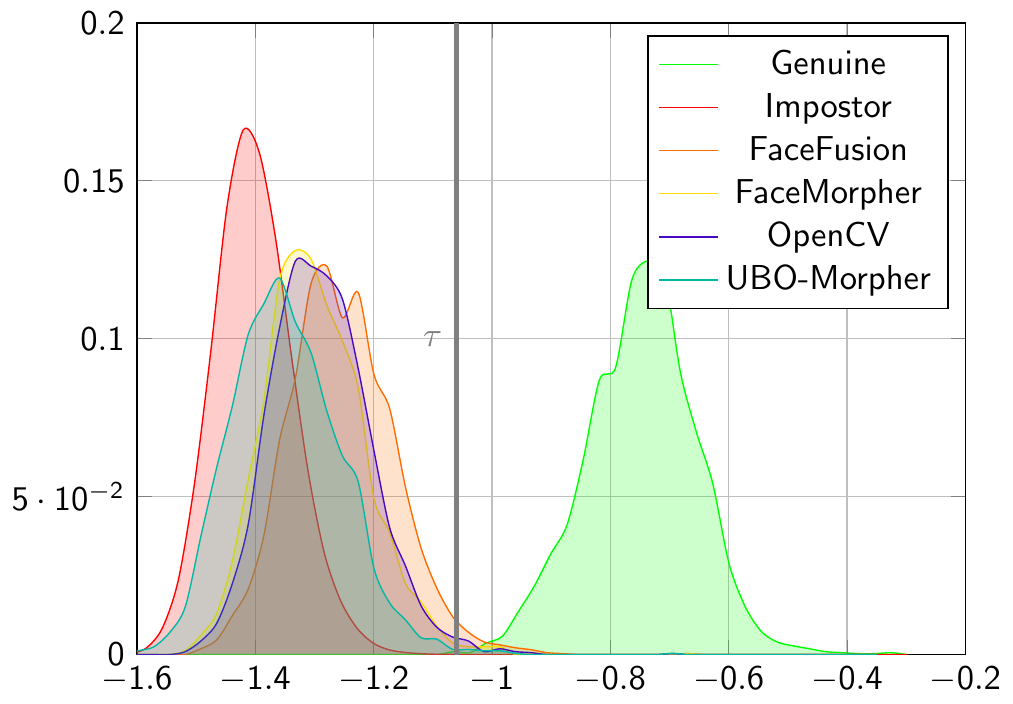}}
	\subfloat[FRGCv2: COTS, MAs$_{25}$]{\includegraphics[height=3.25cm]{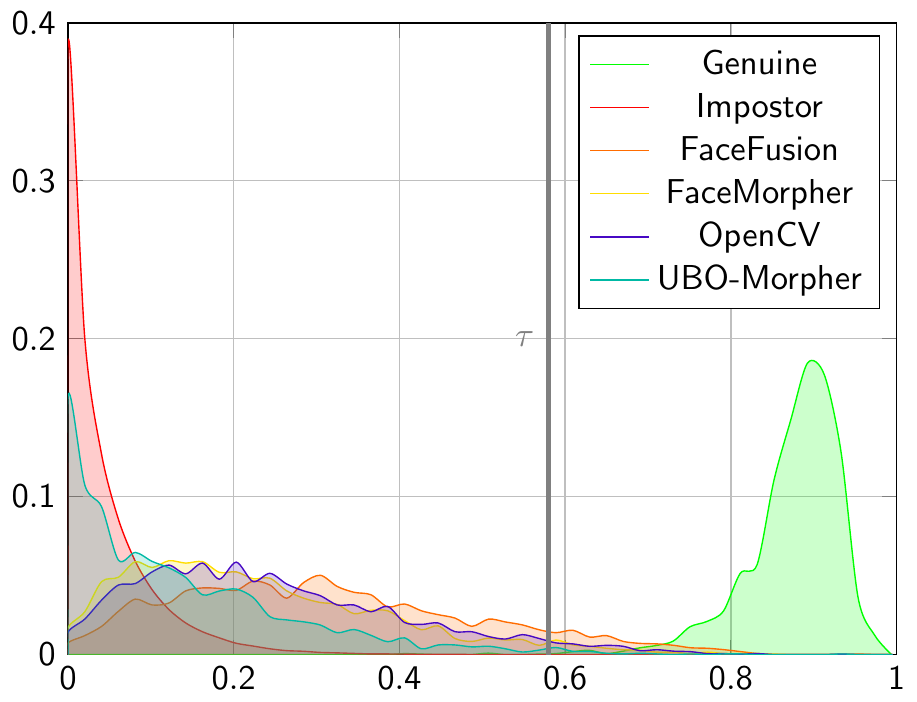}}	
	\caption{Probability Density Functions of comparison scores of genuine, impostor and morph comparisons for all three face recognition systems. Distance scores obtained from ArcFace were multiplied by $-1$ to obtain similarity scores. $\tau$ depicts the estimated threshold for a FMR of 0.1\%.}\label{fig:vuln_pdf}
\end{figure*}

The newly created databases differ considerably from other databases used in scientific publications on MA and MAD. In particular, the intra-class variation is much higher in our database due to the selection of unconstrained probe images. While this approach ensures that our database is more eligible to simulate real-world scenarios, it is perfectly valid to question whether the use of unconstrained probe images may render MA ineffective. Previous analyses of the vulnerability of face recognition systems to MAs, e.g., in~\cite{Ferrara2014,Scherhag2018b}, used probe images, which nearly resembled passport images and, hence, these studies do not apply to the face databases used in this work. Therefore, we evaluated whether face recognition systems are also vulnerable to MAs using our newly created databases.

\begin{table}
	\centering
	\caption{Number of comparisons per test set.}
	\label{tab:comparisons}
	\renewcommand*{\arraystretch}{1.2}
	\begin{tabular}{| l |  c | c | c | c |}
		\hline
		\textbf{Database} & \textbf{Bona Fide} & \textbf{Morph Attacks} & \textbf{Impostors}\\
		\hline
		FERET & 791 & 791 & 418,966 \\
		\hline
		FRGCv2 & 3,298 & 3,246 & 1,695,086\\
		\hline
	\end{tabular} 
\end{table}

\begin{table*}[!t]
	\centering
	\caption{Vulnerability assessment of face recognition systems (FRSs).}
	\label{tab:database_vuln}
	\renewcommand*{\arraystretch}{1.2}
\begin{tabular}{|l|l|c|c|c|c|c|c|c|c|c|c|}
	\cline{5-12}
	 \multicolumn{4}{c|}{} & \multicolumn{8}{c|}{\textbf{MMPMR/RMMR (in \%)}}\\ \hline
	\multirow{2}{*}{\textbf{FRS}} & \multirow{2}{*}{\textbf{Database}} & \multirow{2}{*}{\textbf{\begin{tabular}[c]{@{}c@{}}Decision\\ Threshold \end{tabular}}} & \multirow{2}{*}{\textbf{\begin{tabular}[c]{@{}c@{}}FNMR at\\  FMR=0.1\%\end{tabular}}} & \multicolumn{2}{c|}{\textbf{FaceFusion}} & \multicolumn{2}{c|}{\textbf{FaceMorpher}} & \multicolumn{2}{c|}{\textbf{OpenCV}} & \multicolumn{2}{c|}{\textbf{UBO-Morpher}} \\ \cline{5-12} 
	&                                    &                                                                                             &                                                                                       & \textbf{MAs$_{50}$}      & \textbf{MAs$_{25}$}      & \textbf{MAs$_{50}$}       & \textbf{MAs$_{25}$}      & \textbf{MAs$_{50}$}    & \textbf{MAs$_{25}$}    & \textbf{MAs$_{50}$}   & \textbf{MAs$_{25}$}  \\ \hline
	\multirow{2}{*}{ArcFace}      & FERET                              & -1.10                                                                                       & 0.0                                                                                   & 93.0              & 17.3             & 75.7               & 5.0             & 79.9            & 6.7           & 92.9           & 2.3         \\ \cline{2-12} 
	& FRGCv2                             & -1.06                                                                                       & 0.0                                                                                   & 77.2              & 14.7             & 50.0               & 8.9             & 54.0            & 9.2           & 72.4           & 7.2         \\ \hline
	\multirow{2}{*}{COTS}         & FERET                              & 0.54                                                                                        & 0.0                                                                                   & 96.0              & 33.1             & 88.6               & 10.0             & 91.3            & 13.0           & 96.9           & 5.5         \\ \cline{2-12} 
	& FRGCv2                             & 0.58                                                                                        & 0.0                                                                                   & 79.4              & 18.8             & 60.1               & 8.4             & 62.8            & 9.8           & 81.5           & 3.0         \\ \hline
\end{tabular}
\end{table*}

Firstly, the face recognition performance is evaluated. The amount of bona fide, i.e., genuine, and impostor comparisons, i.e., all possible cross-comparisons of reference and probe images of different subjects, for each database is summarized in Table~\ref{tab:comparisons}. On each database, the decision thresholds of the face recognition systems are set to achieve a False Match Rate (FMR) of $0.1\%$ according to the FRONTEX recommendation for border control scenarios~\cite{FRONTEX2015}. The results are presented in Table~\ref{tab:database_vuln} and the corresponding Probability Density Functions (PDFs) are depicted in Figure~\ref{fig:vuln_pdf}. For ArcFace and the COTS system, no false non-matches occur at an FMR of 0.1\%. Note that this is also to be expected for using color probe images. In general, it can be observed, that the genuine score distributions of FERET are further right (higher similarity scores) than those of FRGCv2. This is likely due to the fact, that the probes of FRGCv2 contain a much higher variation in illumination, sharpness and expression. Further, it can be seen, that the impostor distributions of FERET and FRGCv2 are close to each other, thus the thresholds are approximately the same over both datasets.

Secondly, the vulnerability of the face recognition systems against MAs is evaluated. The vulnerability assessment was conducted using the metrics presented described in \cite{Scherhag2017a}, i.e., Mated Morph Presentation Match Rate (MMPMR) and Relative Morph Match Rate (RMMR). The MMPMR describes the proportion of morphed face images accepted by the face recognition system, the RMMR describes the relation between the MMPMR and the true match rate. Since on both databases no false non-matches occur for the considered decision threshold the MMPMR and the RMMR are equal. The vulnerability analysis is performed exclusively on the NPP image sets of each database, as preliminary studies have shown that the performance of facial recognition systems is only slightly affected by the previously described post-processings. An open-source face recognition system, namely ArcFace \cite{Deng2018}, and one commercial off-the-shelf system, referred to as COTS\footnote{We stress that this COTS system is not Eyedea, which is only used for MAD.}, are employed in the vulnerability assessment. The number of MAs performed for each database is summarized in Table~\ref{tab:comparisons}. The corresponding results are depicted in Figure~\ref{fig:vuln_pdf} and summarized in Table~\ref{tab:database_vuln}.  

Two types of MAs are considered in the vulnerability analysis. On the one hand, morphs with equal weights are employed, i.e., facial images of both subjects contribute with a weight of 50\% to the resulting morph. MAs based on these morphs are denoted as MAs$_{50}$. On the other hand, morphs in which the attacker contributes only with 25\% to the morph are used. Corresponding MAs are referred to as MAs$_{25}$. This means, a larger weight is assigned to an accomplice, i.e., 75\%, who would present the morphed face image to a human observer during the application process of an identity document. Assuming that morphs generated with equal weights of two face images are easily spotted by a human inspection during the application process, the latter type of MAs might be considered as more realistic \cite{Ferrara2018}.    

Focusing on MAs$_{50}$, ArcFace and the COTS system are both very vulnerable to the MAs contained in the database. This can be observed from the high MMPMR/RMMR values in Table~\ref{tab:database_vuln}. Especially on the FERET dataset almost all MAs are accepted. In contrast, the MMPMR/RMMR values, i.e., success chance, significantly drop in case of MAs$_{25}$. 
In general, morphs created by morphing algorithms producing less artefacts (FaceFusion and UBO-Morpher) are more likely to pass the recognition system. In addition, it should be noted that the morphs generated from the FERET database are generally more successful in MAs than the morphs from FRGCv2. This can be attributed to the different intra-class variations of the two databases. Since the genuine comparisons of the FERET data set achieve higher similarity scores, the ones obtained by MAs also tend to be higher. As MAs$_{25}$ achieve only small MMPMR/RMMR values only MAs$_{50}$ are considered for MAD. In case the attacker contributes with a lower weight to the morph compared the accomplice the distance between the resulting morph and the probe image of the attacker increases. Consequently, differential MAD method will generally detect such MAs with higher chance, compared to MAs where morphs have been created using equal weights of the contributing face images.

\section{MAD based on Deep Face Representations}
\label{sec:system}
One of the greatest issue regarding existing differential MAD algorithms is that they can not cope with the large intra-class variance that must be expected for realistic probe images. Deep face recognition networks, however, have shown that they are able to work very robustly, even on challenging data. Therefore, we propose to employ deep face representations extracted by such deep face recognition systems for differential MAD. Precisely, we use the following deep face recognition systems:
\begin{itemize}
 \item ArcFace \cite{Deng2018}, an up-to-date network with a topology optimized for automatic facial recognition
 \item The commercial face recognition SDK from Eyedea\footnote{\url{https://www.eyedea.cz/eyeface-sdk/}}
 \item A re-implementation\footnote{David Sandberg - Face Recognition using Tensorflow, URL: \url{https://github.com/davidsandberg/facenet}} of the FaceNet algorithm \cite{Schroff2015} 
\end{itemize}
  
In principle, it would by possible to apply transfer learning and re-train a pre-trained deep face recognition network to detect morphs. However, the high complexity of the model, represented by the large number of weights in the neural network, requires a large amount of training data. Even if only the lower layers are re-trained, as done in \cite{Ferrara2019}, the limited number of training images (and much lower number of subjects) in our database can easily result in overfitting to the characteristics of the training set. 

\begin{figure}
	\centering
	\includegraphics[width=\linewidth]{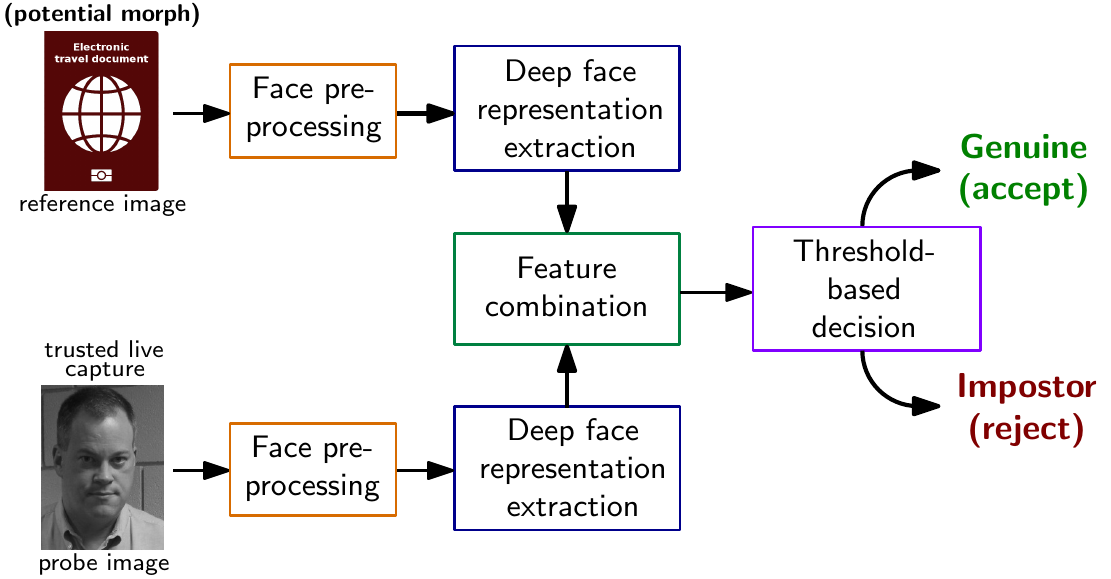}
	\caption{Generic processing chain of face recognition systems based on neural networks.}
	\label{fig:schema_dff_facerec}
\end{figure}
\begin{figure}[!t]
	\centering
	\includegraphics[width=\linewidth]{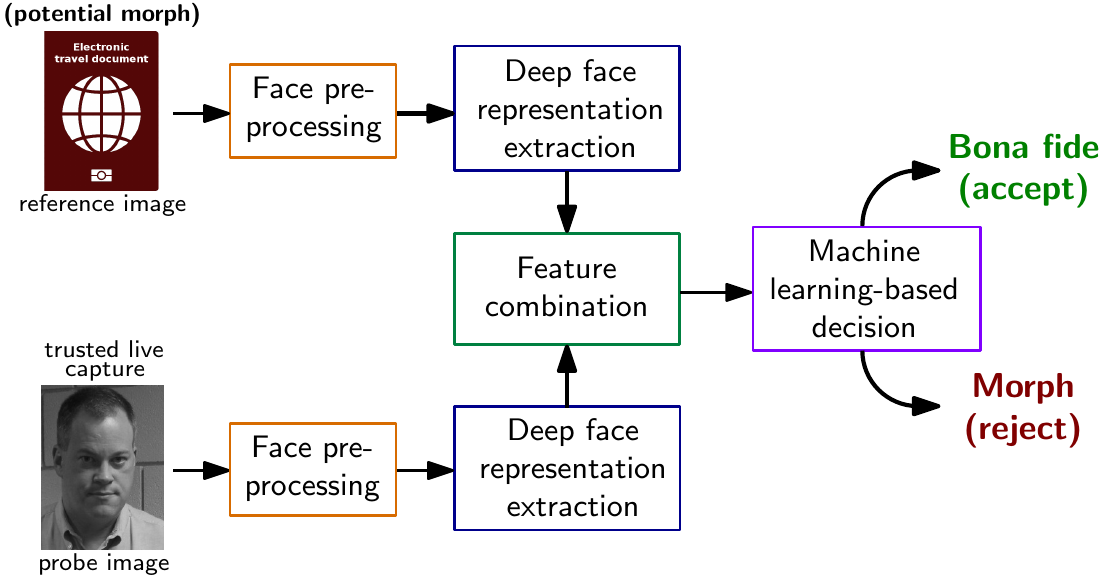}
	\caption{Overview of the proposed differential MAD system based on deep face representations.}
	\label{fig:schema_dff_feat}	
\end{figure}

Therefore, we follow an alternative approach. We use the pre-trained deep face recognition networks as feature extractors and train our MAD algorithms on the deep representations extracted by the neural network (on the lowest layer). Deep face recognition systems leverage very large databases of face images to learn rich and compact representations of faces. While these feature vectors, i.e., deep face representations, have not been trained to detect MAs, at least in the differential scenario, they can, nevertheless, be very useful for MAD: As a morphed face image does not only contain biometric information of the attacker but also those of the accomplice, its deep face representation is expected to, at least in certain aspects, considerably deviate from those detected in the probe image. On the other hand, since there were no morphed facial images in the training set of the neural network, the features can not contain information on image characteristics specific for a certain morphing technique or tool, which reduces the risk of overfitting.

\setlength{\tabcolsep}{3pt}

Most deep face recognition systems work as shown in Figure~\ref{fig:schema_dff_facerec}. The facial image is pre-processed and transferred to the neural network trained for the extraction of deep face representations. This net transfers the facial image into a discriminatory feature space with smaller dimension (512 in the case of ArcFace and FaceNet, 256 in the case of Eyedea). If two images are to be compared, the distance of their feature vectors, e.g., using the Euclidean distance, can serve as dissimilarity score. Even though our vulnerability analysis in Section~\ref{ssec:vulnerability} has shown that this measure is not suitable to separate morphs from bona fide images, the feature vectors can nevertheless contain sufficient information to detect MAs. 

The processing of our MAD algorithms is shown in Figure~\ref{fig:schema_dff_feat}. In the training stage, pre-processed bona fide or morphed reference images and probe images are fed into the neural network. The resulting deep face representations are combined and subsequently processed by a machine learning-based classifier to learn to distinguish between bona fide authentication attempts and MAs. During testing, a potentially morphed reference image and a probe image are processed in the same way and the previously trained classifier is used to estimate the MAD score. A simple but effective combination of the deep features is subtraction, i.e. estimation of a difference vector, which preserves the dimensionality of the feature vector and keeps the training effort low. For the classifier, we tested various machine learning algorithms, namely AdaBoost, Random Forest, Gradient Boosting and SVM. Consistently, SVM with radial basis function as kernel showed the best accuracy and was, thus, chosen for the MAD algorithm.

\begin{figure}[!t]
	\subfloat[FERET: ArcFace]{\includegraphics[width=0.49\linewidth]{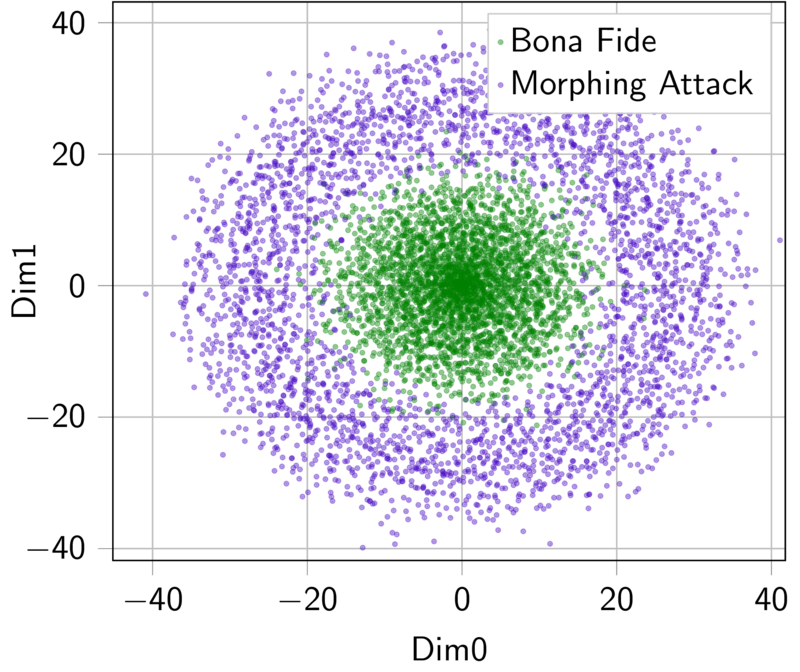}}
	\subfloat[FRGCv2: ArcFace]{\includegraphics[width=0.49\linewidth]{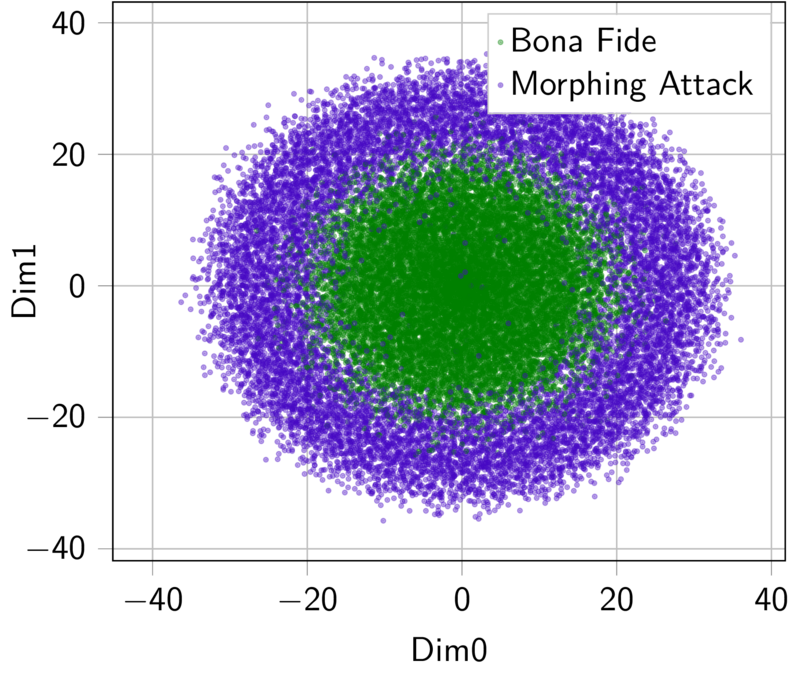}}\\
	\subfloat[FERET: Eyedea]{\includegraphics[width=0.49\linewidth]{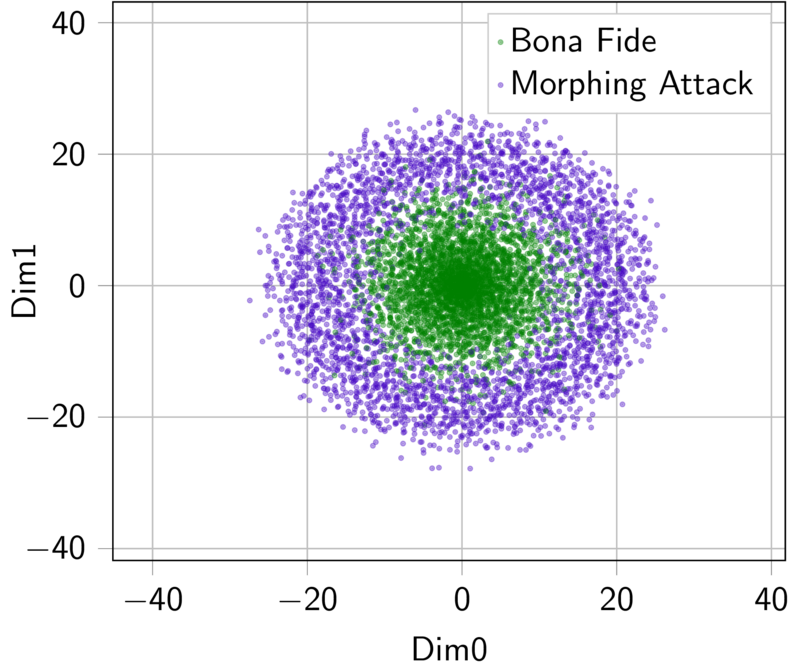}}
	\subfloat[FRGCv2: Eyedea]{\includegraphics[width=0.49\linewidth]{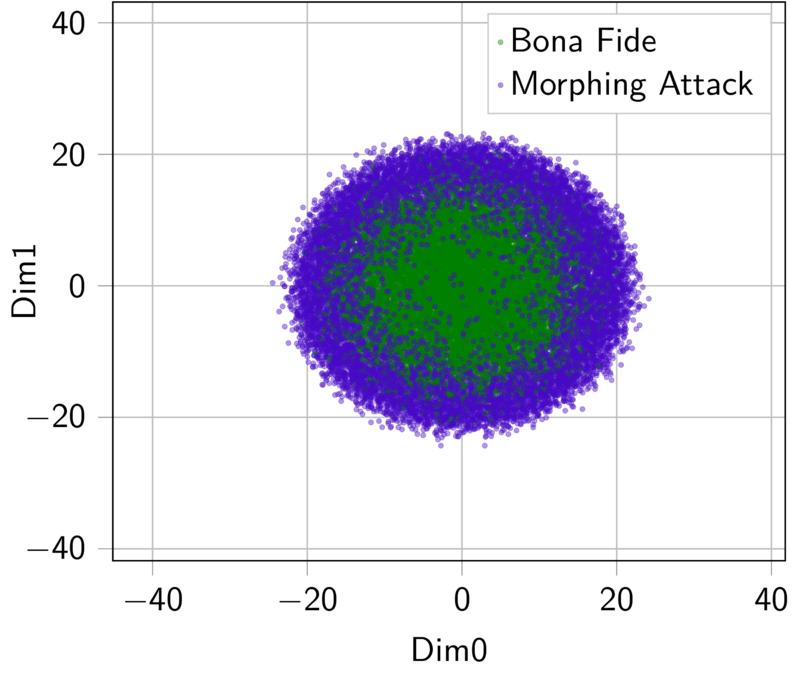}}\\
	\subfloat[FERET: FaceNet]{\includegraphics[width=0.49\linewidth]{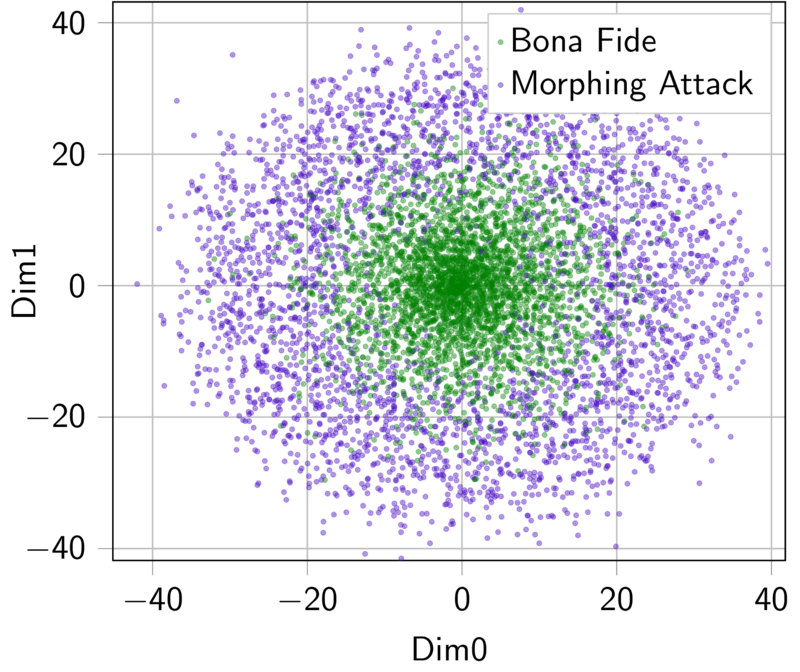}}	
	\subfloat[FRGCv2: FaceNet]{\includegraphics[width=0.49\linewidth]{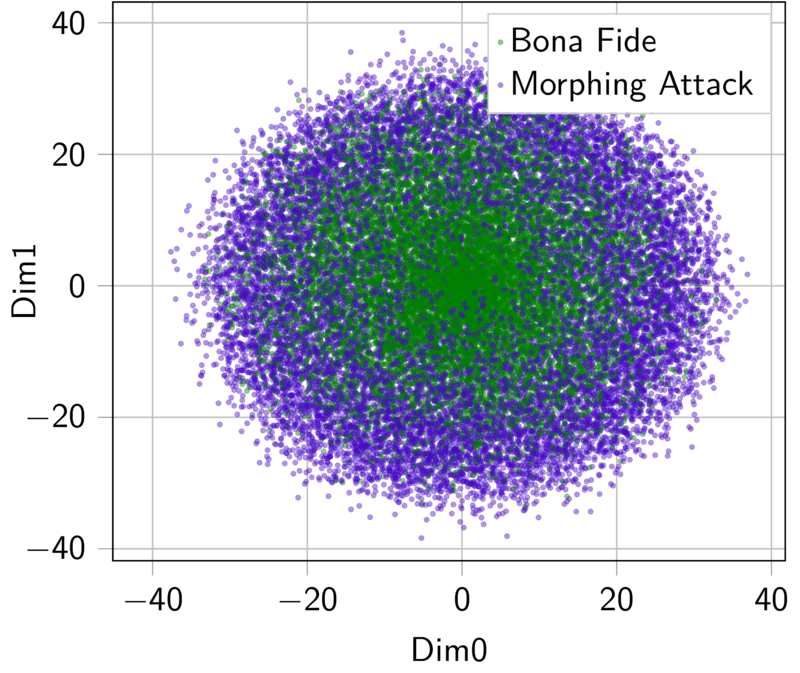}}	
	\caption{Scatter plots of MDS-reduced difference vectors of deep face representations extracted from reference and probe images.}
	\label{fig:mds-plots-frgc}
\end{figure}

By employing the difference vectors of feature vectors extracted from reference and probe images, a simple dimension reduction by Multi-Dimensional Scaling (MDS) to two dimensions shows that bona fide and morphs can be separated. Corresponding scatter plots are provided in Figure~\ref{fig:mds-plots-frgc}. Each plot shows data points for MAs using all of the four morphing tools. Since, as shown in Section~\ref{sec:image_pairing}, the subset of the FRGCv2 allows more comparisons, the corresponding scatter plots are much denser compared to the ones generated from the FERET database. Nevertheless, for both databases it can be seen that MAs and bona fide authentication attempts can be separated on the basis of the difference vectors of deep face representations obtained from ArcFace and Eyedea. Especially the FERET database can be separated almost error-free using the ArcFace algorithm for feature extraction. The FaceNet feature vectors also allow a separation, but higher error rates are expected.

\section{Experiments}
\label{sec:experiments}

Using the databases described in Section~\ref{sec:setup}, we evaluate our MAD approach based on deep face representations described in Section~\ref{sec:system}. In our first experiment, we benchmark the accuracy of our approach together with that of other differential MAD algorithms in the absence of image post-processing. Then, we evaluate the robustness of our approach against image post-processing. Finally, we analyze the score distributions in more detail to gain insight on the causes of classification errors.

For all evaluations presented in this section, we separate training and test sets by source database; precisely, all algorithms are trained using the default hyperparameters on images originating from FERET and evaluated on images from FRGCv2, and vice versa. This approach does not only ensure a strict separation of training and test data, but also a large variance in the image characteristics between these sets. Furthermore, for all evaluations, each of the training sets contains only images with the one of the four post-processings and morphs created with one of the four morphing algorithms (see Section \ref{sec:setup} for details); i.e., we do not combine several post-processings or morphing tools in one training set. The numbers of MAs and bona fide authentication attempts performed for each database during training and testing are listed in Table~\ref{tab:comparisons}.

\begin{table*}[]
		\caption{Detection performance of MAD algorithms on NPP images using different morphing tools (MTs) for training and test.}
	\label{tab:performance-non-pre-processed}
	\scriptsize
	\setlength{\tabcolsep}{2pt}
	\renewcommand*{\arraystretch}{1.2}
\resizebox{\textwidth}{!}{
\begin{tabular}{lll|c|c|c|c|c|c|c|c|c|c|}
\cline{4-13}
&                                                   &                  & \multicolumn{10}{c|}{\textbf{D-EER $|$ BPCER-10 (in \%)}}                                                                                                                                                                                                                                                                          \\ \hline
		\multicolumn{1}{|c|}{\textbf{Train DB}}        & \multicolumn{1}{c|}{\textbf{Train MT}}            & \textbf{Test MT} & \multicolumn{1}{c|}{ArcFace} & \multicolumn{1}{c|}{FaceNet} & \multicolumn{1}{c|}{Eyedea} & \multicolumn{1}{c|}{BSIF} & \multicolumn{1}{c|}{BSIF$_{4\times4}$} & \multicolumn{1}{c|}{LBP} & \multicolumn{1}{c|}{LBP$_{4\times4}$} & \multicolumn{1}{c|}{LM-Wing} & \multicolumn{1}{c|}{LM-Dlib} & Demorphing  \\ \hline
		\multicolumn{1}{|c|}{\multirow{16}{*}{FERET}}  & \multicolumn{1}{c|}{\multirow{4}{*}{FaceFusion}}  & FaceFusion       & \textbf{6.2$\,|\,$3.5}         & 25.9$\,|\,$52.7                & 15.1$\,|\,$22.6               & 39.6$\,|\,$75.8                        & 39.8$\,|\,$77.5                        & 38.7$\,|\,$82.4                       & 41.0$\,|\,$82.0                       & 42.5$\,|\,$83.3                & 44.5$\,|\,$83.0                & 8.6$\,|\,$4.4 \\ \cline{3-13} 
		\multicolumn{1}{|c|}{}                         & \multicolumn{1}{c|}{}                             & FaceMorpher      & \textbf{3.6$\,|\,$0.7}         & 21.7$\,|\,$41.0                & 12.4$\,|\,$14.7               & 41.8$\,|\,$83.5                        & 43.5$\,|\,$86.5                        & 38.6$\,|\,$91.6                       & 42.7$\,|\,$92.7                       & 41.5$\,|\,$81.2                & 43.6$\,|\,$83.0                & 3.4$\,|\,$0.0 \\ \cline{3-13} 
		\multicolumn{1}{|c|}{}                         & \multicolumn{1}{c|}{}                             & OpenCV           & \textbf{3.7$\,|\,$0.8}         & 21.7$\,|\,$40.7                & 12.3$\,|\,$14.8               & 42.8$\,|\,$82.8                        & 43.5$\,|\,$83.7                        & 39.7$\,|\,$86.5                       & 42.6$\,|\,$87.9                       & 42.5$\,|\,$82.1                & 43.9$\,|\,$82.8                & 3.8$\,|\,$0.8 \\ \cline{3-13} 
		\multicolumn{1}{|c|}{}                         & \multicolumn{1}{c|}{}                             & UBO-Morpher              & \textbf{6.0$\,|\,$3.5}         & 25.5$\,|\,$50.5                & 16.2$\,|\,$24.5               & 42.5$\,|\,$81.4                        & 43.3$\,|\,$83.0                        & 39.2$\,|\,$85.1                       & 42.8$\,|\,$85.3                       & 42.3$\,|\,$82.3                & 44.4$\,|\,$83.3                & 7.8$\,|\,$5.8 \\ \cline{2-13} 
		\multicolumn{1}{|c|}{}                         & \multicolumn{1}{c|}{\multirow{4}{*}{FaceMorpher}} & FaceFusion       & \textbf{7.1$\,|\,$4.5}         & 26.3$\,|\,$53.4                & 16.2$\,|\,$24.9               & 49.3$\,|\,$83.3                        & 46.7$\,|\,$83.4                        & 48.1$\,|\,$87.1                       & 46.3$\,|\,$87.5                       & 43.9$\,|\,$81.7                & 43.8$\,|\,$88.1                & 8.6$\,|\,$4.4 \\ \cline{3-13} 
		\multicolumn{1}{|c|}{}                         & \multicolumn{1}{c|}{}                             & FaceMorpher      & \textbf{3.3$\,|\,$0.5}         & 19.8$\,|\,$36.4                & 10.3$\,|\,$10.4               & 50.2$\,|\,$88.0                        & 47.5$\,|\,$88.6                        & 48.1$\,|\,$93.2                       & 45.5$\,|\,$95.2                       & 42.0$\,|\,$79.1                & 43.0$\,|\,$87.7                & 3.4$\,|\,$0.0 \\ \cline{3-13} 
		\multicolumn{1}{|c|}{}                         & \multicolumn{1}{c|}{}                             & OpenCV           & \textbf{3.7$\,|\,$0.7}         & 20.7$\,|\,$37.8                & 10.5$\,|\,$11.5               & 51.3$\,|\,$86.5                        & 49.1$\,|\,$86.5                        & 48.8$\,|\,$89.0                       & 47.1$\,|\,$90.7                       & 43.2$\,|\,$80.0                & 42.7$\,|\,$87.2                & 3.8$\,|\,$0.8 \\ \cline{3-13} 
		\multicolumn{1}{|c|}{}                         & \multicolumn{1}{c|}{}                             & UBO-Morpher              & \textbf{6.6$\,|\,$4.2}         & 25.2$\,|\,$49.5                & 17.2$\,|\,$26.3               & 51.1$\,|\,$86.5                        & 48.7$\,|\,$85.3                        & 48.1$\,|\,$88.8                       & 47.4$\,|\,$89.1                       & 43.7$\,|\,$81.6                & 42.9$\,|\,$87.1                & 7.8$\,|\,$5.8 \\ \cline{2-13} 
		\multicolumn{1}{|c|}{}                         & \multicolumn{1}{c|}{\multirow{4}{*}{OpenCV}}      & FaceFusion       & \textbf{6.8$\,|\,$4.0}         & 26.8$\,|\,$53.6                & 16.3$\,|\,$23.9               & 46.9$\,|\,$82.0                        & 43.3$\,|\,$80.7                        & 44.9$\,|\,$86.0                       & 43.7$\,|\,$84.8                       & 43.1$\,|\,$83.2                & 45.5$\,|\,$84.2                & 8.6$\,|\,$4.4 \\ \cline{3-13} 
		\multicolumn{1}{|c|}{}                         & \multicolumn{1}{c|}{}                             & FaceMorpher      & \textbf{3.2$\,|\,$0.7}         & 20.1$\,|\,$36.2                & 11.0$\,|\,$12.0               & 46.4$\,|\,$84.9                        & 44.9$\,|\,$87.1                        & 42.8$\,|\,$92.9                       & 43.2$\,|\,$94.8                       & 41.2$\,|\,$80.9                & 44.2$\,|\,$84.0                & 3.4$\,|\,$0.0 \\ \cline{3-13} 
		\multicolumn{1}{|c|}{}                         & \multicolumn{1}{c|}{}                             & OpenCV           & \textbf{3.3$\,|\,$0.8}         & 20.1$\,|\,$35.8                & 10.7$\,|\,$11.9               & 48.2$\,|\,$84.7                        & 45.8$\,|\,$85.2                        & 44.9$\,|\,$88.9                       & 44.3$\,|\,$88.9                       & 42.0$\,|\,$81.5                & 44.4$\,|\,$83.7                & 3.8$\,|\,$0.8 \\ \cline{3-13} 
		\multicolumn{1}{|c|}{}                         & \multicolumn{1}{c|}{}                             & UBO              & \textbf{6.3$\,|\,$3.6}         & 24.9$\,|\,$49.9                & 16.8$\,|\,$24.2               & 48.3$\,|\,$84.2                        & 45.3$\,|\,$84.0                        & 44.8$\,|\,$88.5                       & 44.8$\,|\,$87.1                       & 42.8$\,|\,$83.0                & 45.1$\,|\,$84.6                & 7.8$\,|\,$5.8 \\ \cline{2-13} 
		\multicolumn{1}{|c|}{}                         & \multicolumn{1}{c|}{\multirow{4}{*}{UBO-Morpher}}         & FaceFusion       & \textbf{6.7$\,|\,$4.2}         & 26.6$\,|\,$54.2                & 16.4$\,|\,$24.2               & 47.0$\,|\,$81.0                        & 44.4$\,|\,$81.7                        & 43.7$\,|\,$84.2                       & 44.0$\,|\,$85.8                       & 42.7$\,|\,$82.5                & 44.8$\,|\,$84.3                & 8.6$\,|\,$4.4 \\ \cline{3-13} 
		\multicolumn{1}{|c|}{}                         & \multicolumn{1}{c|}{}                             & FaceMorpher      & \textbf{3.8$\,|\,$1.1}         & 21.8$\,|\,$42.3                & 12.8$\,|\,$17.1               & 45.7$\,|\,$84.7                        & 46.7$\,|\,$87.9                        & 39.7$\,|\,$91.2                       & 44.2$\,|\,$94.1                       & 41.5$\,|\,$80.9                & 44.2$\,|\,$86.0                & 3.4$\,|\,$0.0 \\ \cline{3-13} 
		\multicolumn{1}{|c|}{}                         & \multicolumn{1}{c|}{}                             & OpenCV           & \textbf{3.8$\,|\,$1.2}         & 21.6$\,|\,$39.3                & 12.7$\,|\,$16.5               & 47.1$\,|\,$84.5                        & 46.7$\,|\,$85.4                        & 42.6$\,|\,$87.1                       & 44.2$\,|\,$89.5                       & 42.5$\,|\,$81.5                & 44.4$\,|\,$84.2                & 3.8$\,|\,$0.8 \\ \cline{3-13} 
		\multicolumn{1}{|c|}{}                         & \multicolumn{1}{c|}{}                             & UBO-Morpher              & \textbf{5.7$\,|\,$3.0}         & 24.4$\,|\,$48.7                & 15.5$\,|\,$22.7               & 47.2$\,|\,$82.8                        & 45.6$\,|\,$83.5                        & 41.6$\,|\,$85.4                       & 43.0$\,|\,$85.6                       & 41.6$\,|\,$82.0                & 44.3$\,|\,$82.4                & 7.8$\,|\,$5.8 \\ \hline
		\multicolumn{1}{|c|}{\multirow{16}{*}{FRGCv2}} & \multicolumn{1}{c|}{\multirow{4}{*}{FaceFusion}}  & FaceFusion       & \textbf{2.1$\,|\,$0.6}         & 13.0$\,|\,$15.4                & 5.3$\,|\,$3.1                 & 15.5$\,|\,$23.8                        & 15.9$\,|\,$28.3                        & 18.1$\,|\,$30.2                       & 25.9$\,|\,$49.1                       & 39.4$\,|\,$84.4                & 38.9$\,|\,$75.7                & 6.7$\,|\,$5.1 \\ \cline{3-13} 
		\multicolumn{1}{|c|}{}                         & \multicolumn{1}{c|}{}                             & FaceMorpher      & \textbf{1.1$\,|\,$0.0}         & 9.6$\,|\,$9.4                  & 4.3$\,|\,$1.9                 & 14.5$\,|\,$18.8                        & 13.8$\,|\,$24.6                        & 12.9$\,|\,$17.4                       & 21.9$\,|\,$38.1                       & 41.2$\,|\,$85.5                & 37.6$\,|\,$71.7                & 4.1$\,|\,$1.3 \\ \cline{3-13} 
		\multicolumn{1}{|c|}{}                         & \multicolumn{1}{c|}{}                             & OpenCV           & \textbf{1.4$\,|\,$0.0}         & 9.8$\,|\,$9.4                  & 4.0$\,|\,$2.4                 & 15.9$\,|\,$27.0                        & 16.7$\,|\,$30.2                        & 16.9$\,|\,$28.2                       & 24.3$\,|\,$47.9                       & 41.3$\,|\,$87.0                & 39.4$\,|\,$75.8                & 4.2$\,|\,$1.4 \\ \cline{3-13} 
		\multicolumn{1}{|c|}{}                         & \multicolumn{1}{c|}{}                             & UBO-Morpher              & \textbf{2.5$\,|\,$0.9}         & 13.0$\,|\,$15.4                & 6.9$\,|\,$5.1                 & 16.3$\,|\,$25.5                        & 18.1$\,|\,$32.4                        & 17.0$\,|\,$27.2                       & 24.3$\,|\,$45.3                       & 40.2$\,|\,$86.2                & 39.5$\,|\,$77.5                & 7.7$\,|\,$6.1 \\ \cline{2-13} 
		\multicolumn{1}{|c|}{}                         & \multicolumn{1}{c|}{\multirow{4}{*}{FaceMorpher}} & FaceFusion       & \textbf{2.7$\,|\,$1.1}         & 14.4$\,|\,$18.1                & 7.3$\,|\,$5.7                 & 24.9$\,|\,$47.5                        & 23.1$\,|\,$52.0                        & 33.0$\,|\,$63.9                       & 37.7$\,|\,$75.7                       & 40.6$\,|\,$82.5                & 39.8$\,|\,$78.6                & 6.7$\,|\,$5.1 \\ \cline{3-13} 
		\multicolumn{1}{|c|}{}                         & \multicolumn{1}{c|}{}                             & FaceMorpher      & \textbf{0.9$\,|\,$0.0}         & 8.4$\,|\,$6.5                  & 3.1$\,|\,$1.2                 & 15.1$\,|\,$21.7                        & 15.5$\,|\,$24.3                        & 24.6$\,|\,$46.2                       & 31.7$\,|\,$63.1                       & 40.8$\,|\,$83.5                & 39.8$\,|\,$83.7                & 4.1$\,|\,$1.3 \\ \cline{3-13} 
		\multicolumn{1}{|c|}{}                         & \multicolumn{1}{c|}{}                             & OpenCV           & \textbf{1.1$\,|\,$0.0}         & 8.2$\,|\,$6.8                  & 3.9$\,|\,$1.6                 & 19.3$\,|\,$33.0                        & 19.3$\,|\,$40.1                        & 30.2$\,|\,$57.2                       & 34.6$\,|\,$69.6                       & 41.9$\,|\,$84.8                & 40.8$\,|\,$81.9                & 4.2$\,|\,$1.4 \\ \cline{3-13} 
		\multicolumn{1}{|c|}{}                         & \multicolumn{1}{c|}{}                             & UBO-Morpher              & \textbf{2.5$\,|\,$1.1}         & 13.6$\,|\,$16.0                & 8.3$\,|\,$7.3                 & 21.6$\,|\,$36.4                        & 20.7$\,|\,$44.0                        & 30.3$\,|\,$57.5                       & 35.4$\,|\,$70.3                       & 41.0$\,|\,$83.2                & 42.1$\,|\,$87.0                & 7.7$\,|\,$6.1 \\ \cline{2-13} 
		\multicolumn{1}{|c|}{}                         & \multicolumn{1}{c|}{\multirow{4}{*}{OpenCV}}      & FaceFusion       & \textbf{2.7$\,|\,$0.9}         & 14.0$\,|\,$19.2                & 7.3$\,|\,$5.3                 & 18.4$\,|\,$31.9                        & 19.3$\,|\,$34.0                        & 23.9$\,|\,$44.8                       & 30.7$\,|\,$62.4                       & 40.8$\,|\,$83.2                & 38.2$\,|\,$75.2                & 6.7$\,|\,$5.1 \\ \cline{3-13} 
		\multicolumn{1}{|c|}{}                         & \multicolumn{1}{c|}{}                             & FaceMorpher      & \textbf{0.6$\,|\,$0.0}         & 9.3$\,|\,$8.4                  & 2.9$\,|\,$1.1                 & 14.1$\,|\,$18.3                        & 15.3$\,|\,$23.9                        & 18.6$\,|\,$33.4                       & 25.5$\,|\,$51.3                       & 41.3$\,|\,$85.9                & 35.6$\,|\,$73.1                & 4.1$\,|\,$1.3 \\ \cline{3-13} 
		\multicolumn{1}{|c|}{}                         & \multicolumn{1}{c|}{}                             & OpenCV           & \textbf{0.9$\,|\,$0.0}         & 8.9$\,|\,$7.9                  & 3.4$\,|\,$1.2                 & 15.4$\,|\,$25.3                        & 18.1$\,|\,$30.6                        & 21.9$\,|\,$43.0                       & 28.8$\,|\,$58.7                       & 41.2$\,|\,$86.3                & 38.0$\,|\,$75.3                & 4.2$\,|\,$1.4 \\ \cline{3-13} 
		\multicolumn{1}{|c|}{}                         & \multicolumn{1}{c|}{}                             & UBO-Morpher              & \textbf{2.7$\,|\,$0.9}         & 13.1$\,|\,$16.9                & 7.5$\,|\,$6.7                 & 17.4$\,|\,$26.5                        & 18.1$\,|\,$32.4                        & 21.7$\,|\,$42.7                       & 29.3$\,|\,$57.0                       & 41.9$\,|\,$85.8                & 38.8$\,|\,$77.7                & 7.7$\,|\,$6.1 \\ \cline{2-13} 
		\multicolumn{1}{|c|}{}                         & \multicolumn{1}{c|}{\multirow{4}{*}{UBO-Morpher}}         & FaceFusion       & \textbf{3.2$\,|\,$0.7}         & 13.4$\,|\,$16.7                & 6.0$\,|\,$4.0                 & 20.5$\,|\,$33.9                        & 18.8$\,|\,$35.0                        & 21.7$\,|\,$46.7                       & 31.0$\,|\,$61.0                       & 41.1$\,|\,$86.3                & 38.4$\,|\,$76.6                & 6.7$\,|\,$5.1 \\ \cline{3-13} 
		\multicolumn{1}{|c|}{}                         & \multicolumn{1}{c|}{}                             & FaceMorpher      & \textbf{0.7$\,|\,$0.0}         & 9.8$\,|\,$9.6                  & 4.4$\,|\,$2.4                 & 13.6$\,|\,$19.6                        & 14.5$\,|\,$22.2                        & 16.8$\,|\,$25.5                       & 25.4$\,|\,$49.7                       & 43.0$\,|\,$86.7                & 36.0$\,|\,$71.3                & 4.1$\,|\,$1.3 \\ \cline{3-13} 
		\multicolumn{1}{|c|}{}                         & \multicolumn{1}{c|}{}                             & OpenCV           & \textbf{1.1$\,|\,$0.0}         & 9.4$\,|\,$9.1                  & 4.1$\,|\,$2.4                 & 16.4$\,|\,$24.3                        & 18.2$\,|\,$28.3                        & 20.2$\,|\,$41.6                       & 28.8$\,|\,$58.6                       & 42.5$\,|\,$88.4                & 39.4$\,|\,$74.7                & 4.2$\,|\,$1.4 \\ \cline{3-13} 
		\multicolumn{1}{|c|}{}                         & \multicolumn{1}{c|}{}                             & UBO-Morpher              & \textbf{2.4$\,|\,$0.7}         & 11.5$\,|\,$13.6                & 5.4$\,|\,$4.0                 & 17.5$\,|\,$23.9                        & 18.2$\,|\,$27.8                        & 19.8$\,|\,$38.8                       & 28.4$\,|\,$56.5                       & 40.8$\,|\,$87.2                & 38.7$\,|\,$75.5                & 7.7$\,|\,$6.1 \\ \hline
\end{tabular}
}
\end{table*}

The accuracy of the detection algorithms is reported using the Detection Equal Error Rate (D-EER), i.e., at the decision threshold where the proportion of attack presentations incorrectly classified as bona fide presentations (APCER) is as high as the proportion of bona fide presentations incorrectly classified as presentation attack (BPCER). For APCER and BPCER the definitions of ISO IEC 30107-3 \cite{ISOIECJTCSCB2017} for measuring accuracy of presentation attack detection are used:
\begin{quote}
	\textit{APCER}: proportion of attack presentations incorrectly classified as bona fide presentations in a specific scenario
\end{quote}
\begin{quote}
	\textit{BPCER}: proportion of bona fide presentations incorrectly classified as presentation attacks in a specific scenario
\end{quote}
Additionally, the BPCER10 is reported, i.e.,  the operation  point  where  APCER=10\%.

\subsection{Detection Accuracy in the Absence of Post-Processing and Comparison with other MAD Algorithms}

\begin{table*}[]
	\scriptsize
	\caption{Detection performance of MAD algorithms based on deep face representations with different post-processings (PPs).}
	\label{tab:performance-with-pre-processing}
	\renewcommand*{\arraystretch}{1.2}
	\setlength{\tabcolsep}{1.7pt}
	\centering
\begin{tabular}{cccl|c|c|c|c|c|c|c|c|c|c|c|c|}
\cline{5-16}
&                                                   &                                                   &                  & \multicolumn{12}{c|}{\textbf{D-EER $|$ BPCER-10 (in \%)}}                                                                                                                                \\ \cline{5-16} 
		&                                                   &                                                   &                  & NPP         & Resized     & JP2         & PS-JP2      & NPP           & Resized       & JP2           & PS-JP2        & NPP           & Resized       & JP2           & PS-JP2        \\ \hline
		\multicolumn{1}{|c|}{\textbf{Train DB}}       & \multicolumn{1}{c|}{\textbf{Train MA}}            & \multicolumn{1}{c|}{\textbf{Test MA}}             & \textbf{Test PP} & \multicolumn{4}{c|}{ArcFace}                          & \multicolumn{4}{c|}{Eyedea}                                   & \multicolumn{4}{c|}{FaceNet}                                  \\ \hline
		\multicolumn{1}{|c|}{\multirow{16}{*}{FERET}} & \multicolumn{1}{c|}{\multirow{8}{*}{FaceMorpher}} & \multicolumn{1}{c|}{\multirow{4}{*}{FaceFusion}}  & NPP              & 7.3$\,|\,$4.6 & 7.2$\,|\,$4.6 & 7.1$\,|\,$5.0 & 7.3$\,|\,$5.3 & 16.2$\,|\,$24.6 & 16.2$\,|\,$24.9 & 16.1$\,|\,$25.3 & 16.5$\,|\,$25.5 & 27.5$\,|\,$54.5 & 27.5$\,|\,$54.3 & 27.5$\,|\,$55.4 & 28.1$\,|\,$56.8 \\ \cline{4-16} 
		\multicolumn{1}{|c|}{}                        & \multicolumn{1}{c|}{}                             & \multicolumn{1}{c|}{}                             & Resized          & 7.1$\,|\,$4.5 & 7.1$\,|\,$4.6 & 7.1$\,|\,$5.0 & 7.3$\,|\,$5.1 & 16.2$\,|\,$24.9 & 16.2$\,|\,$25.1 & 16.3$\,|\,$24.9 & 16.7$\,|\,$25.8 & 26.3$\,|\,$53.4 & 25.8$\,|\,$53.4 & 26.3$\,|\,$52.6 & 26.7$\,|\,$56.3 \\ \cline{4-16} 
		\multicolumn{1}{|c|}{}                        & \multicolumn{1}{c|}{}                             & \multicolumn{1}{c|}{}                             & JP2              & 7.7$\,|\,$5.9 & 7.7$\,|\,$6.0 & 7.4$\,|\,$6.2 & 7.7$\,|\,$6.2 & 16.8$\,|\,$26.7 & 16.9$\,|\,$26.1 & 16.5$\,|\,$26.1 & 16.8$\,|\,$26.0 & 26.9$\,|\,$54.1 & 26.8$\,|\,$54.2 & 26.8$\,|\,$53.8 & 27.8$\,|\,$55.5 \\ \cline{4-16} 
		\multicolumn{1}{|c|}{}                        & \multicolumn{1}{c|}{}                             & \multicolumn{1}{c|}{}                             & PS-JP2           & 7.2$\,|\,$4.6 & 7.1$\,|\,$4.5 & 7.2$\,|\,$5.0 & 7.3$\,|\,$5.2 & 16.2$\,|\,$24.3 & 16.1$\,|\,$24.5 & 15.9$\,|\,$24.3 & 16.4$\,|\,$24.5 & 27.3$\,|\,$54.5 & 27.0$\,|\,$54.9 & 27.6$\,|\,$55.1 & 27.9$\,|\,$57.4 \\ \cline{3-16} 
		\multicolumn{1}{|c|}{}                        & \multicolumn{1}{c|}{}                             & \multicolumn{1}{c|}{\multirow{4}{*}{UBO-Morpher}} & NPP              & 6.6$\,|\,$4.2 & 6.6$\,|\,$4.2 & 6.6$\,|\,$4.4 & 7.0$\,|\,$4.5 & 17.2$\,|\,$27.4 & 17.2$\,|\,$26.6 & 17.4$\,|\,$26.9 & 17.1$\,|\,$26.8 & 25.5$\,|\,$51.8 & 25.7$\,|\,$51.4 & 25.8$\,|\,$51.6 & 26.1$\,|\,$54.3 \\ \cline{4-16} 
		\multicolumn{1}{|c|}{}                        & \multicolumn{1}{c|}{}                             & \multicolumn{1}{c|}{}                             & Resized          & 6.6$\,|\,$4.2 & 6.7$\,|\,$4.1 & 6.6$\,|\,$4.4 & 6.8$\,|\,$4.4 & 17.2$\,|\,$26.3 & 17.3$\,|\,$26.2 & 17.2$\,|\,$25.9 & 17.1$\,|\,$26.3 & 25.2$\,|\,$49.5 & 24.7$\,|\,$49.0 & 24.9$\,|\,$49.5 & 25.8$\,|\,$51.2 \\ \cline{4-16} 
		\multicolumn{1}{|c|}{}                        & \multicolumn{1}{c|}{}                             & \multicolumn{1}{c|}{}                             & JP2              & 7.0$\,|\,$5.0 & 7.0$\,|\,$4.9 & 6.9$\,|\,$4.9 & 6.7$\,|\,$4.9 & 17.6$\,|\,$28.3 & 17.4$\,|\,$27.9 & 17.7$\,|\,$27.7 & 17.4$\,|\,$27.7 & 25.8$\,|\,$53.9 & 26.1$\,|\,$51.9 & 26.1$\,|\,$52.6 & 26.7$\,|\,$53.6 \\ \cline{4-16} 
		\multicolumn{1}{|c|}{}                        & \multicolumn{1}{c|}{}                             & \multicolumn{1}{c|}{}                             & PS-JP2           & 6.7$\,|\,$4.2 & 6.7$\,|\,$4.2 & 6.5$\,|\,$4.3 & 6.8$\,|\,$4.6 & 17.0$\,|\,$27.1 & 17.2$\,|\,$26.8 & 17.0$\,|\,$26.3 & 17.0$\,|\,$26.6 & 25.6$\,|\,$51.5 & 25.2$\,|\,$51.9 & 25.6$\,|\,$51.9 & 26.1$\,|\,$53.6 \\ \cline{2-16} 
		\multicolumn{1}{|c|}{}                        & \multicolumn{1}{c|}{\multirow{8}{*}{OpenCV}}      & \multicolumn{1}{c|}{\multirow{4}{*}{FaceFusion}}  & NPP              & 6.6$\,|\,$4.0 & 6.6$\,|\,$3.7 & 6.7$\,|\,$4.0 & 6.8$\,|\,$4.3 & 16.2$\,|\,$24.6 & 16.2$\,|\,$24.1 & 16.3$\,|\,$25.2 & 16.4$\,|\,$25.3 & 27.1$\,|\,$54.8 & 26.8$\,|\,$54.8 & 26.7$\,|\,$54.9 & 27.3$\,|\,$56.2 \\ \cline{4-16} 
		\multicolumn{1}{|c|}{}                        & \multicolumn{1}{c|}{}                             & \multicolumn{1}{c|}{}                             & Resized          & 6.8$\,|\,$4.0 & 6.7$\,|\,$4.1 & 6.7$\,|\,$4.4 & 6.9$\,|\,$4.7 & 16.3$\,|\,$23.9 & 16.1$\,|\,$23.7 & 16.4$\,|\,$24.8 & 16.7$\,|\,$25.7 & 26.8$\,|\,$53.6 & 25.5$\,|\,$52.8 & 25.8$\,|\,$52.5 & 26.5$\,|\,$54.6 \\ \cline{4-16} 
		\multicolumn{1}{|c|}{}                        & \multicolumn{1}{c|}{}                             & \multicolumn{1}{c|}{}                             & JP2              & 7.2$\,|\,$4.9 & 7.0$\,|\,$4.9 & 7.0$\,|\,$5.2 & 7.0$\,|\,$5.2 & 16.8$\,|\,$25.6 & 16.6$\,|\,$25.4 & 16.8$\,|\,$26.2 & 16.5$\,|\,$26.4 & 26.9$\,|\,$53.4 & 26.1$\,|\,$53.1 & 26.4$\,|\,$53.9 & 27.0$\,|\,$55.0 \\ \cline{4-16} 
		\multicolumn{1}{|c|}{}                        & \multicolumn{1}{c|}{}                             & \multicolumn{1}{c|}{}                             & PS-JP2           & 6.7$\,|\,$4.2 & 6.6$\,|\,$4.2 & 6.7$\,|\,$4.3 & 7.0$\,|\,$4.5 & 16.3$\,|\,$24.3 & 15.9$\,|\,$24.2 & 16.3$\,|\,$24.6 & 16.3$\,|\,$24.8 & 27.2$\,|\,$54.6 & 26.2$\,|\,$54.7 & 26.1$\,|\,$55.0 & 26.9$\,|\,$56.2 \\ \cline{3-16} 
		\multicolumn{1}{|c|}{}                        & \multicolumn{1}{c|}{}                             & \multicolumn{1}{c|}{\multirow{4}{*}{UBO-Morpher}} & NPP              & 6.1$\,|\,$3.5 & 6.0$\,|\,$3.4 & 6.2$\,|\,$3.7 & 6.4$\,|\,$4.0 & 16.6$\,|\,$25.7 & 16.8$\,|\,$25.1 & 16.9$\,|\,$26.5 & 16.6$\,|\,$26.0 & 25.7$\,|\,$51.4 & 25.1$\,|\,$50.4 & 25.0$\,|\,$49.4 & 25.4$\,|\,$52.8 \\ \cline{4-16} 
		\multicolumn{1}{|c|}{}                        & \multicolumn{1}{c|}{}                             & \multicolumn{1}{c|}{}                             & Resized          & 6.3$\,|\,$3.6 & 6.3$\,|\,$3.5 & 6.4$\,|\,$3.8 & 6.5$\,|\,$4.0 & 16.8$\,|\,$24.2 & 17.0$\,|\,$24.0 & 17.0$\,|\,$25.3 & 16.8$\,|\,$25.6 & 24.9$\,|\,$49.9 & 24.1$\,|\,$48.6 & 24.1$\,|\,$48.9 & 25.1$\,|\,$50.1 \\ \cline{4-16} 
		\multicolumn{1}{|c|}{}                        & \multicolumn{1}{c|}{}                             & \multicolumn{1}{c|}{}                             & JP2              & 6.7$\,|\,$4.0 & 6.5$\,|\,$4.1 & 6.5$\,|\,$3.9 & 6.4$\,|\,$4.0 & 17.3$\,|\,$26.4 & 17.3$\,|\,$26.0 & 17.4$\,|\,$27.7 & 16.9$\,|\,$26.6 & 25.8$\,|\,$52.7 & 25.7$\,|\,$52.1 & 25.6$\,|\,$52.0 & 25.7$\,|\,$52.9 \\ \cline{4-16} 
		\multicolumn{1}{|c|}{}                        & \multicolumn{1}{c|}{}                             & \multicolumn{1}{c|}{}                             & PS-JP2           & 6.3$\,|\,$3.7 & 6.2$\,|\,$3.8 & 6.2$\,|\,$4.0 & 6.5$\,|\,$4.2 & 17.0$\,|\,$25.5 & 16.6$\,|\,$25.2 & 17.2$\,|\,$26.1 & 16.6$\,|\,$26.3 & 25.7$\,|\,$50.2 & 24.7$\,|\,$49.3 & 24.6$\,|\,$49.4 & 25.3$\,|\,$52.0 \\ \hline
		\multicolumn{1}{|c|}{\multirow{16}{*}{FRGCv2}}  & \multicolumn{1}{c|}{\multirow{8}{*}{FaceMorpher}} & \multicolumn{1}{c|}{\multirow{4}{*}{FaceFusion}}  & NPP              & 2.9$\,|\,$1.2 & 3.2$\,|\,$1.1 & 2.7$\,|\,$1.0 & 2.8$\,|\,$1.1 & 7.5$\,|\,$6.3   & 7.5$\,|\,$6.5   & 7.3$\,|\,$6.0   & 7.4$\,|\,$5.9   & 14.1$\,|\,$18.4 & 14.5$\,|\,$19.3 & 14.4$\,|\,$19.3 & 14.9$\,|\,$21.2 \\ \cline{4-16} 
		\multicolumn{1}{|c|}{}                        & \multicolumn{1}{c|}{}                             & \multicolumn{1}{c|}{}                             & Resized          & 2.7$\,|\,$1.1 & 2.9$\,|\,$1.2 & 2.7$\,|\,$1.2 & 2.8$\,|\,$1.2 & 7.3$\,|\,$5.7   & 7.4$\,|\,$5.9   & 7.0$\,|\,$5.9   & 7.0$\,|\,$5.7   & 14.4$\,|\,$18.1 & 15.3$\,|\,$18.8 & 14.4$\,|\,$18.3 & 15.0$\,|\,$18.8 \\ \cline{4-16} 
		\multicolumn{1}{|c|}{}                        & \multicolumn{1}{c|}{}                             & \multicolumn{1}{c|}{}                             & JP2              & 1.2$\,|\,$0.1 & 1.2$\,|\,$0.1 & 1.2$\,|\,$0.1 & 1.2$\,|\,$0.1 & 8.3$\,|\,$6.8   & 8.4$\,|\,$6.9   & 8.1$\,|\,$6.8   & 7.5$\,|\,$7.3   & 10.0$\,|\,$10.3 & 10.2$\,|\,$10.6 & 10.6$\,|\,$11.0 & 11.0$\,|\,$12.1 \\ \cline{4-16} 
		\multicolumn{1}{|c|}{}                        & \multicolumn{1}{c|}{}                             & \multicolumn{1}{c|}{}                             & PS-JP2           & 2.8$\,|\,$1.0 & 3.0$\,|\,$1.0 & 3.0$\,|\,$1.0 & 2.9$\,|\,$1.0 & 7.4$\,|\,$6.0   & 7.8$\,|\,$6.7   & 7.9$\,|\,$6.5   & 7.4$\,|\,$6.2   & 14.4$\,|\,$19.2 & 15.3$\,|\,$20.7 & 15.0$\,|\,$20.2 & 15.8$\,|\,$21.1 \\ \cline{3-16} 
		\multicolumn{1}{|c|}{}                        & \multicolumn{1}{c|}{}                             & \multicolumn{1}{c|}{\multirow{4}{*}{UBO-Morpher}} & NPP              & 3.0$\,|\,$1.2 & 3.2$\,|\,$1.1 & 2.8$\,|\,$1.0 & 3.0$\,|\,$1.1 & 8.2$\,|\,$7.2   & 8.2$\,|\,$7.4   & 8.2$\,|\,$7.0   & 8.3$\,|\,$7.5   & 13.5$\,|\,$16.5 & 14.0$\,|\,$16.4 & 13.8$\,|\,$16.8 & 14.5$\,|\,$18.1 \\ \cline{4-16} 
		\multicolumn{1}{|c|}{}                        & \multicolumn{1}{c|}{}                             & \multicolumn{1}{c|}{}                             & Resized          & 2.5$\,|\,$1.1 & 2.8$\,|\,$1.2 & 2.8$\,|\,$1.2 & 2.8$\,|\,$1.2 & 8.3$\,|\,$7.3   & 8.1$\,|\,$7.3   & 7.7$\,|\,$6.8   & 8.1$\,|\,$6.8   & 13.6$\,|\,$16.0 & 13.6$\,|\,$17.3 & 13.6$\,|\,$16.7 & 14.6$\,|\,$18.8 \\ \cline{4-16} 
		\multicolumn{1}{|c|}{}                        & \multicolumn{1}{c|}{}                             & \multicolumn{1}{c|}{}                             & JP2              & 3.2$\,|\,$1.2 & 3.3$\,|\,$1.2 & 3.7$\,|\,$1.2 & 3.4$\,|\,$1.4 & 9.7$\,|\,$8.2   & 9.4$\,|\,$8.6   & 8.8$\,|\,$7.9   & 8.9$\,|\,$8.1   & 15.3$\,|\,$19.6 & 15.1$\,|\,$20.6 & 14.5$\,|\,$20.2 & 15.3$\,|\,$19.7 \\ \cline{4-16} 
		\multicolumn{1}{|c|}{}                        & \multicolumn{1}{c|}{}                             & \multicolumn{1}{c|}{}                             & PS-JP2           & 2.8$\,|\,$1.0 & 2.8$\,|\,$1.0 & 2.7$\,|\,$1.0 & 3.0$\,|\,$1.0 & 8.3$\,|\,$7.4   & 7.9$\,|\,$7.4   & 8.1$\,|\,$7.3   & 8.1$\,|\,$7.0   & 14.1$\,|\,$18.2 & 14.1$\,|\,$19.3 & 14.5$\,|\,$18.3 & 14.8$\,|\,$19.6 \\ \cline{2-16} 
		\multicolumn{1}{|c|}{}                        & \multicolumn{1}{c|}{\multirow{8}{*}{OpenCV}}      & \multicolumn{1}{c|}{\multirow{4}{*}{FaceFusion}}  & NPP              & 2.7$\,|\,$0.7 & 2.9$\,|\,$0.9 & 2.8$\,|\,$1.0 & 2.8$\,|\,$1.0 & 7.3$\,|\,$5.3   & 7.3$\,|\,$5.3   & 7.5$\,|\,$5.4   & 7.2$\,|\,$5.5   & 14.1$\,|\,$18.2 & 13.6$\,|\,$17.8 & 13.8$\,|\,$18.2 & 15.0$\,|\,$20.3 \\ \cline{4-16} 
		\multicolumn{1}{|c|}{}                        & \multicolumn{1}{c|}{}                             & \multicolumn{1}{c|}{}                             & Resized          & 2.7$\,|\,$0.9 & 2.7$\,|\,$0.9 & 2.5$\,|\,$1.0 & 2.0$\,|\,$1.0 & 7.3$\,|\,$5.3   & 7.2$\,|\,$5.3   & 7.2$\,|\,$5.4   & 6.8$\,|\,$5.5   & 14.0$\,|\,$19.2 & 14.9$\,|\,$17.8 & 15.7$\,|\,$18.4 & 15.1$\,|\,$19.7 \\ \cline{4-16} 
		\multicolumn{1}{|c|}{}                        & \multicolumn{1}{c|}{}                             & \multicolumn{1}{c|}{}                             & JP2              & 1.4$\,|\,$0.1 & 1.4$\,|\,$0.1 & 1.4$\,|\,$0.1 & 1.4$\,|\,$0.1 & 7.5$\,|\,$6.8   & 7.8$\,|\,$6.4   & 7.7$\,|\,$6.7   & 8.9$\,|\,$7.3   & 9.6$\,|\,$9.2   & 9.8$\,|\,$9.8   & 10.0$\,|\,$9.8  & 10.5$\,|\,$11.3 \\ \cline{4-16} 
		\multicolumn{1}{|c|}{}                        & \multicolumn{1}{c|}{}                             & \multicolumn{1}{c|}{}                             & PS-JP2           & 2.9$\,|\,$0.9 & 2.9$\,|\,$0.9 & 2.5$\,|\,$0.7 & 2.4$\,|\,$1.0 & 7.2$\,|\,$5.3   & 7.0$\,|\,$5.7   & 7.2$\,|\,$5.5   & 7.3$\,|\,$5.3   & 14.9$\,|\,$20.0 & 14.0$\,|\,$18.6 & 14.1$\,|\,$19.8 & 15.5$\,|\,$21.3 \\ \cline{3-16} 
		\multicolumn{1}{|c|}{}                        & \multicolumn{1}{c|}{}                             & \multicolumn{1}{c|}{\multirow{4}{*}{UBO-Morpher}} & NPP              & 2.7$\,|\,$0.7 & 2.8$\,|\,$0.7 & 2.7$\,|\,$0.7 & 2.8$\,|\,$0.9 & 7.5$\,|\,$6.3   & 7.5$\,|\,$6.0   & 7.7$\,|\,$6.5   & 7.9$\,|\,$6.5   & 13.1$\,|\,$16.5 & 12.7$\,|\,$15.7 & 13.5$\,|\,$16.3 & 14.3$\,|\,$17.7 \\ \cline{4-16} 
		\multicolumn{1}{|c|}{}                        & \multicolumn{1}{c|}{}                             & \multicolumn{1}{c|}{}                             & Resized          & 2.7$\,|\,$0.9 & 2.8$\,|\,$0.9 & 2.3$\,|\,$0.9 & 2.7$\,|\,$0.9 & 7.5$\,|\,$6.7   & 7.9$\,|\,$6.4   & 7.9$\,|\,$6.7   & 7.9$\,|\,$6.0   & 13.1$\,|\,$16.9 & 13.6$\,|\,$16.9 & 13.4$\,|\,$16.5 & 14.8$\,|\,$18.3 \\ \cline{4-16} 
		\multicolumn{1}{|c|}{}                        & \multicolumn{1}{c|}{}                             & \multicolumn{1}{c|}{}                             & JP2              & 3.0$\,|\,$1.2 & 2.9$\,|\,$1.0 & 2.9$\,|\,$1.0 & 3.2$\,|\,$1.1 & 8.6$\,|\,$7.4   & 8.4$\,|\,$7.2   & 8.4$\,|\,$7.7   & 9.1$\,|\,$8.7   & 14.1$\,|\,$19.6 & 14.3$\,|\,$18.2 & 14.4$\,|\,$19.3 & 14.3$\,|\,$21.3 \\ \cline{4-16} 
		\multicolumn{1}{|c|}{}                        & \multicolumn{1}{c|}{}                             & \multicolumn{1}{c|}{}                             & PS-JP2           & 2.5$\,|\,$0.9 & 2.7$\,|\,$0.9 & 2.5$\,|\,$0.7 & 2.7$\,|\,$0.9 & 7.7$\,|\,$6.4   & 7.9$\,|\,$6.7   & 8.1$\,|\,$6.4   & 7.5$\,|\,$6.3   & 13.5$\,|\,$19.4 & 13.0$\,|\,$17.7 & 13.6$\,|\,$18.1 & 14.8$\,|\,$18.4 \\ \hline
\end{tabular}
\end{table*}

In this subsection, we evaluate the accuracy of our MAD approach on  the NPP images and compare its detection accuracy to that of several other differential MAD approaches. Beside the classifiers trained on deep face representations extracted using ArcFace, Eyedea and FaceNet, we test two MAD algorithms based on texture descriptors, namely LBP \cite{Ojala1996} and BSIF \cite{Kannala2012} with patches of size 3$\times$3 and an optional division into 4$\times$4 cells. In order to adapt these feature extractors to differential detection scenario, the difference between the histograms of reference and probe images are used as classifier input as proposed in~\cite{Scherhag2018b}. Further, two landmark-based algorithms implemented according to \cite{Damer2018b} are evaluated using landmarks computed with Dlib \cite{King2009} and Wing Loss \cite{Feng2017}, respectively. Finally, we evaluate the demorphing algorithm of \cite{Ferrara2018} with a demorphing factor of 0.3 in combination with Dlib landmarks and the COTS face recognition algorithm used in Section~\ref{ssec:vulnerability}. 

The results are shown in Table~\ref{tab:performance-non-pre-processed}. Note that the demorphing algorithm is not trained, which is why for this MAD algorithm in the table the specification of the morphing algorithm used for training have no effect.   Consistently, when testing on FRGCv2 (and thus training on FERET), the detection performance of all MAD algorithms is much lower than that achieved on FERET. This observation can be explained by the fact that, as visible in Figure~\ref{fig:example_probes_frgc}, the probes of the FRGCv2 show a significantly higher variance in illumination, background and sharpness, whereas the probes of the FERET database contain pose variants, but are consistently good in quality, see Figure~\ref{fig:example_probes_feret}. Similarly, our results on the MAD based on texture descriptors and landmarks are much worse than those reported in~\cite{Scherhag2018b} and \cite{Damer2018b}, where probe images of higher quality were used. Thus, we conclude that the quality of the probes has a strong effect on the detection performance of the MAD algorithms. This further underlines the need for databases containing probe images with realistic characteristics. 

Our MAD algorithms based on deep face representations yields superior results compared to the majority of other methods. The algorithm based on ArcFace features by far outperforms all other approaches, achieving very low D-EER between 1\% and 7\%. The algorithm using Eyedea features ranks second with an D-EER between 3\% and 17\%. Obviously, the features of FaceNet are far less suitable for MAD. A clear correlation to the scatter plots shown in Figure~\ref{fig:mds-plots-frgc} can be observed.

The texture-based MAD algorithms are only capable of detecting morphed face images (although with very high error rates) if the probe image is available in sufficient quality, i.e., on the FERET database. The detection performance of these algorithms when probes with a more realistic variance, i.e., on the FRGCv2 database, is close to random. The performance of landmark-based MAD is less dependent on the quality of probe images but generally very poor. In contrast, demorphing in combination with the COTS face recognition system yields good detection rates. In particular, for FRGCv2, it performs only slightly worse than our ArcFace-based MAD for the higher quality morphs (FaceFusion and UBO-Morpher) and even comparably good for the lower quality morphs. However, on the FERET database, its performance even falls behind that of the Eyedea-based MAD algorithm.

Since the use of grayscale probe images is not expected to significantly impact the recognition performance of face recognition systems based on deep face representations it is valid to assume that the proposed MAD approaches achieve equal detection performance in case colour probe images are processed. The same holds for MAD algorithms which explicitly perform a grayscale conversion prior to the feature extraction, e.g., MAD based on LBP or BSIF feature extractors.

A general pattern observable in the results is that morphs with higher quality are more difficult to detect. FaceFusion and the UBO-Morpher both include automatic post-processing, which replaces the artefact-rich region outside the face and therefore generate a higher error rate during MAD.\footnote{In addition, the outer region is replaced with that from the attacker's image which is expected to increase the similarity in comparison with a probe image from the attacker.} Although the best results are achieved if for training and test sets the same morphing tool is used (which indicates slight overfitting to the characteristics of the morphing process), the influence of the morphing tool used for training is quite limited. 

\subsection{Robustness against Image Post-Processing}

In this subsection, the robustness of MAD algorithms based on deep face representations with respect to post-processing of the reference images is investigated. All four post-processings presented in Section~\ref{sec:pre-processing} are separately used for training and testing. Since we have already seen that the morphing tool used for training is of minor importance and that higher quality morphs (created with FaceFusion and UBO-Morpher) are more difficult to detect, we use FaceMorpher and OpenCV for training, and FaceFusion and UBO-Morpher for testing. The other MAD algorithms considered in the previous section will not be further considered, because they cannot achieve comparable performance. 

The corresponding results are shown in Table \ref{tab:performance-with-pre-processing}. As in the previous experiment, the MAD algorithm based on deep face represnetations extracted using ArcFace is far ahead in detection performance, followed by the MAD algorithm based on Eyedea. Training on post-processed images has no noticeable influence on the detection performance of the algorithms. Testing on post-processed images can have a slight effect on detection performance. In general, the proposed MAD algorithms are not severely influenced by the tested post-processings. Presumably, this is due to the extraction of the deep face representations by the neural networks which were trained for a high independence from image properties. 

\begin{figure*}[!t]
	\subfloat[FRGCv2: NPP]{\includegraphics[width=0.25\textwidth]{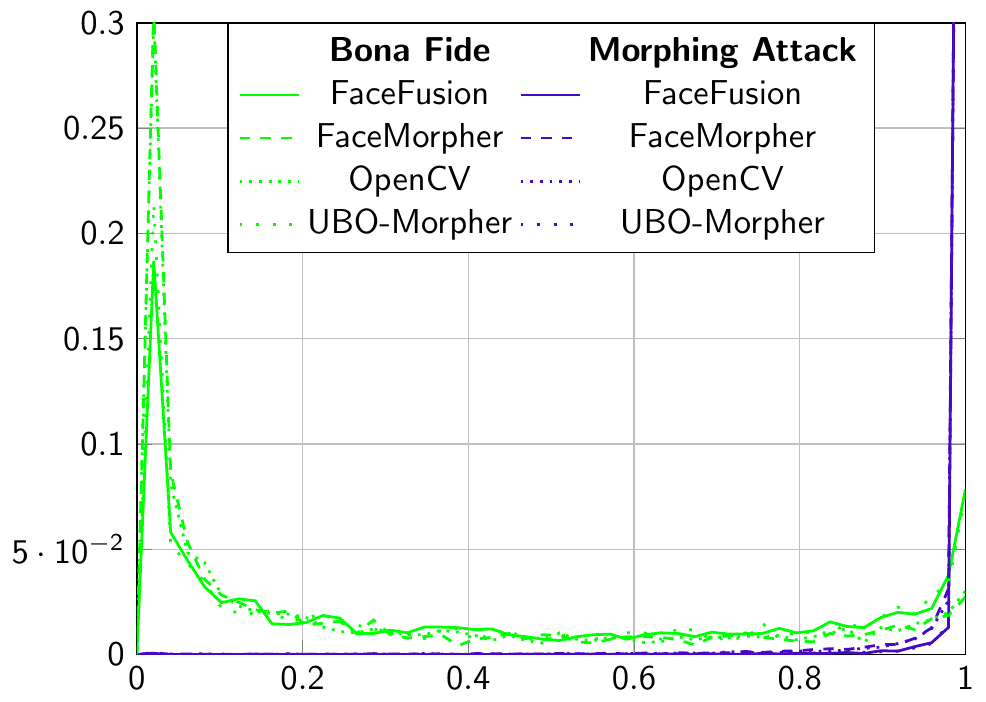}}
	\subfloat[FRGCv2: Resized]{\includegraphics[width=0.25\textwidth]{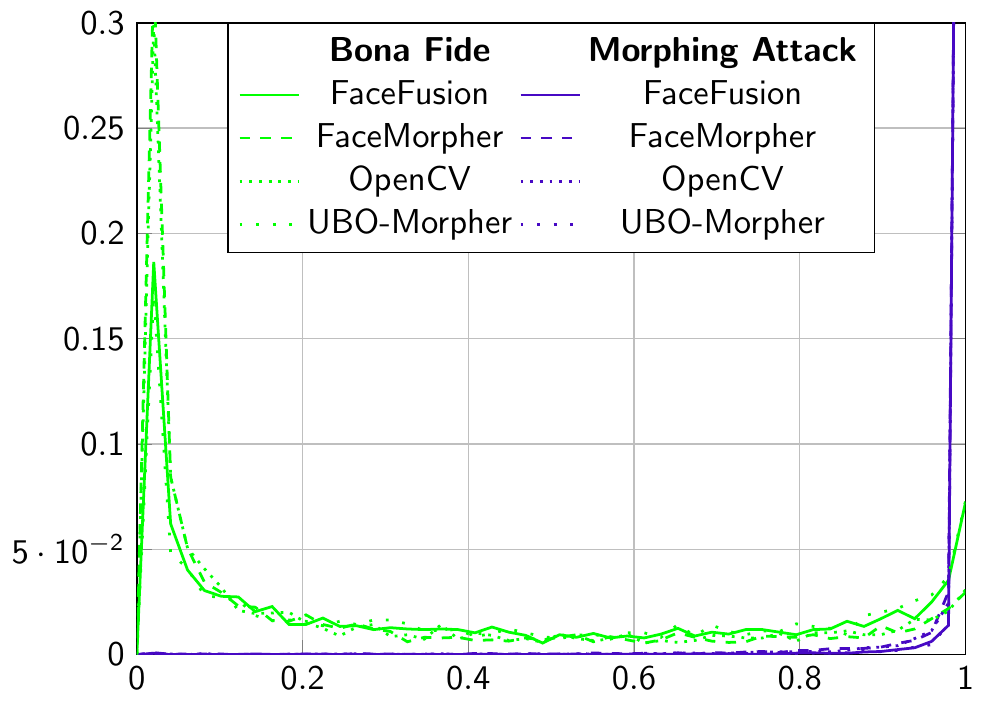}}
	\subfloat[FRGCv2: JP2]{\includegraphics[width=0.25\textwidth]{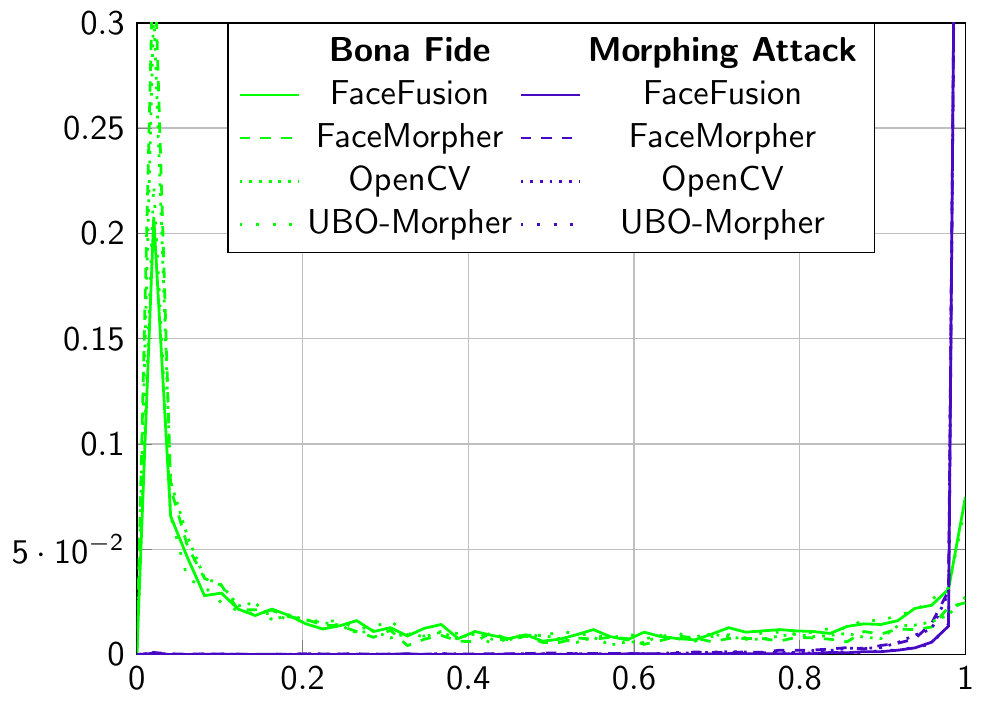}}	
	\subfloat[FRGCv2: PS-JP2]{\includegraphics[width=0.25\textwidth]{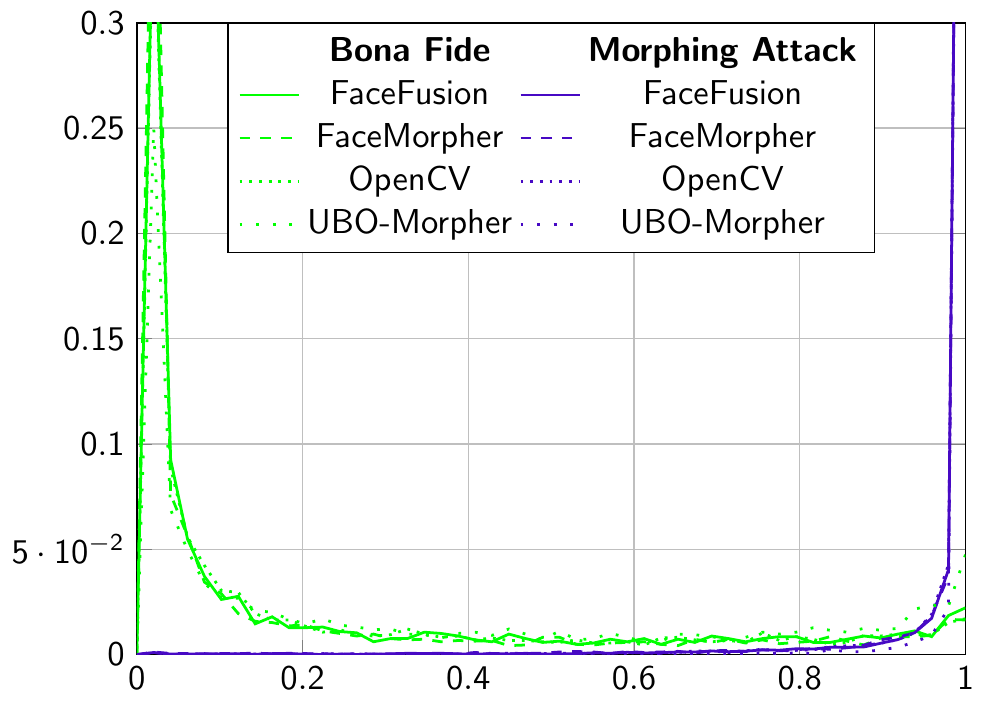}}\\
	\subfloat[FERET: NPP]{\includegraphics[width=0.25\textwidth]{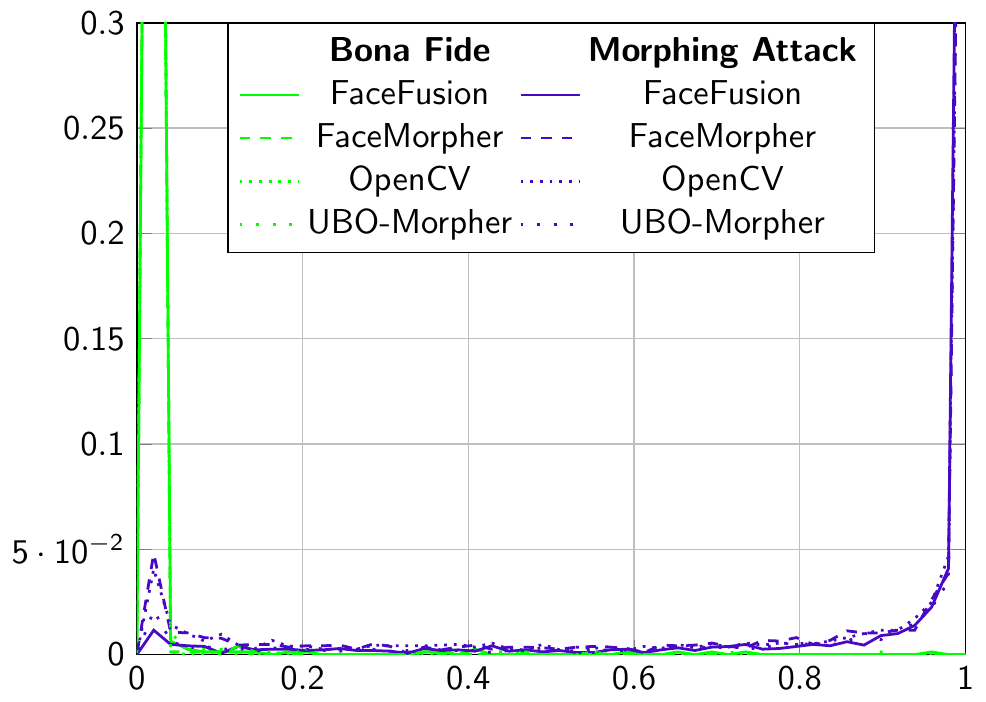}}
	\subfloat[FERET: Resized]{\includegraphics[width=0.25\textwidth]{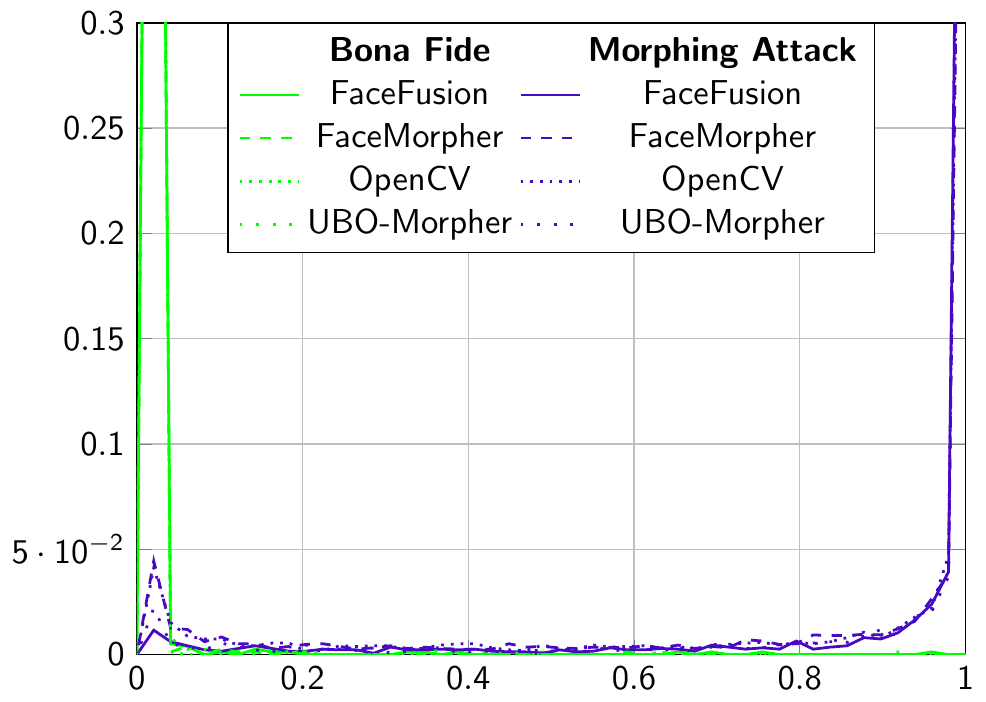}}
	\subfloat[FERET: JP2]{\includegraphics[width=0.25\textwidth]{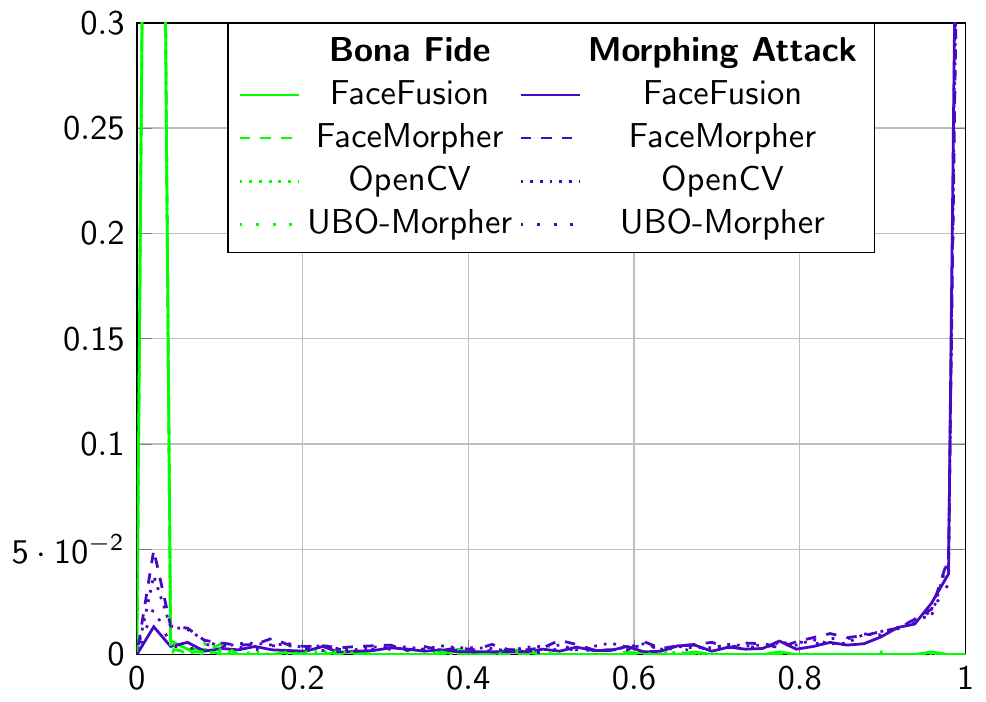}}	
	\subfloat[FERET: PS-JP2]{\includegraphics[width=0.25\textwidth]{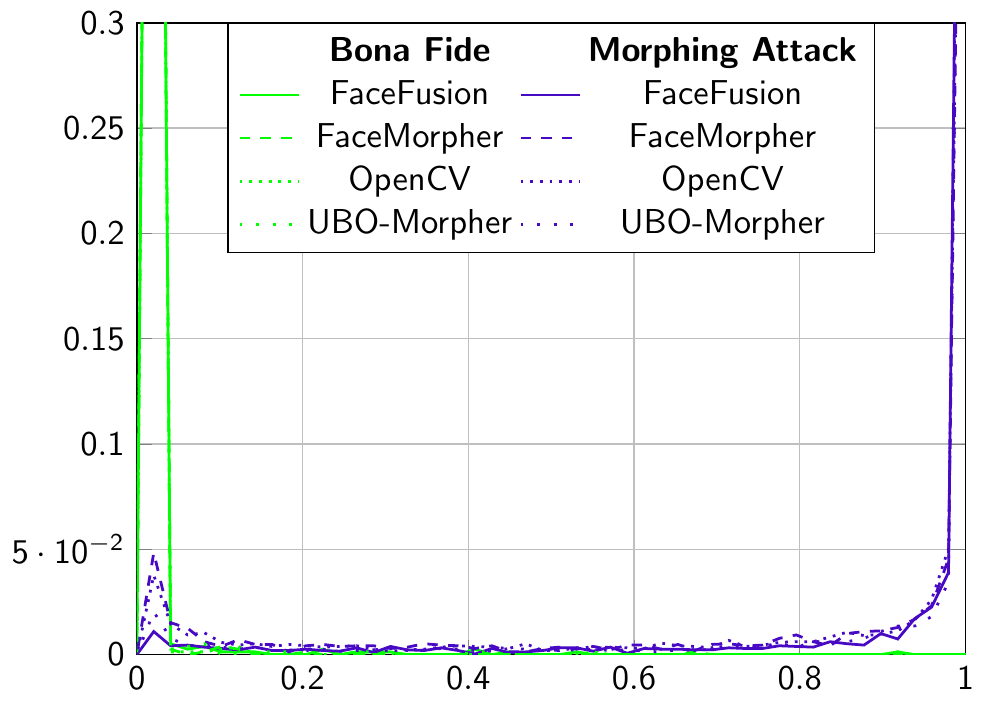}}
	\caption{PDFs of the decision score of an MAD based on ArcFace features for different databases and post-processings, when trained on different morphing attack algorithms. Low scores (close to 0) indicate to bona fide authentication attempts and high scores (close to 1) indicate MAs.}
	\label{fig:performance-pdf}
\end{figure*}

\subsection{Analysis of Score Distributions}

As a final investigation, score distributions of the MAD algorithm based on the ArcFace feature extraction are analyzed in more detail. Since this MAD algorithm achieves by far the best performance, the other two deep face representation extractors (Eyedea and Facenet) are not considered here. Since post-processing has no influence during training, only algorithms trained on NPP images are considered. Trained SVMs generate a normalized MAD score in the range $[0,1]$ where low scores, i.e., scores close to 0, indicate bona fide authentication attempts and high scores, i.e., scores close to 1, refer to MAs. Figure~\ref{fig:performance-pdf} shows the resulting probability density functions. For each subfigure, training was performed with each morphing tool separately (the tool name is given in the legend of the plot), while testing was performed considering MAs using all morphing algorithms together. The database and post-processing used for the evaluation are given in the caption of the respective subfigure. 

The observations correspond to the findings of the previous experiments. Within a database, the score distributions are very similar, regardless of which morphing tool is used for training and which post-processing is used for the test set. This further suggests that the use of a variety of different morphing tools in the training stage is not expected to yield significant improvement in terms of detection performance. Generally, scores which were generated by the evaluation on FERET can be separated better than those stemming from FRGCv2.

\begin{figure}
	\hfill
	\subfloat[Reference 1]{\includegraphics[height = 2.8cm]{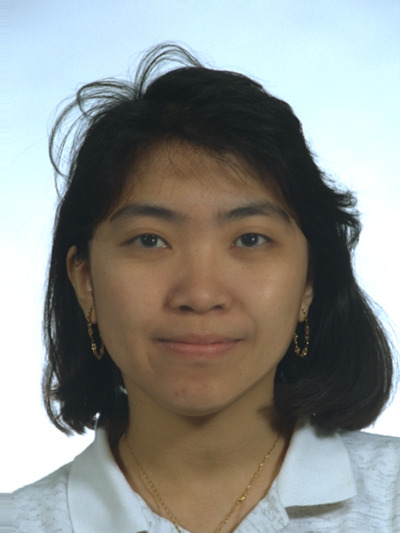}}\hfill
	\subfloat[Probe 1]{\includegraphics[height = 2.8cm]{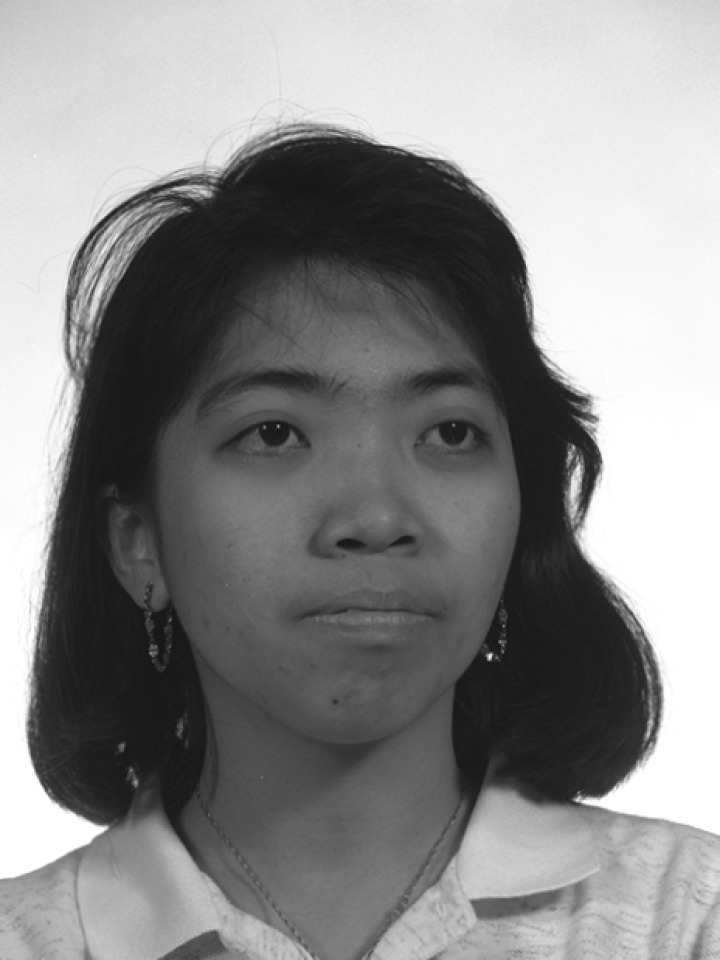}}\hfill
	\subfloat[Reference 2]{\includegraphics[height = 2.8cm]{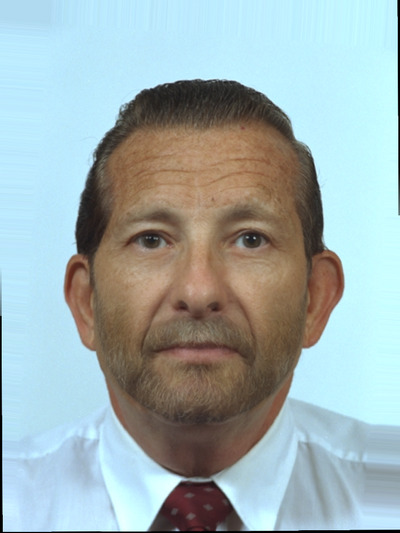}}\hfill
	\subfloat[Probe 2]{\includegraphics[height = 2.8cm]{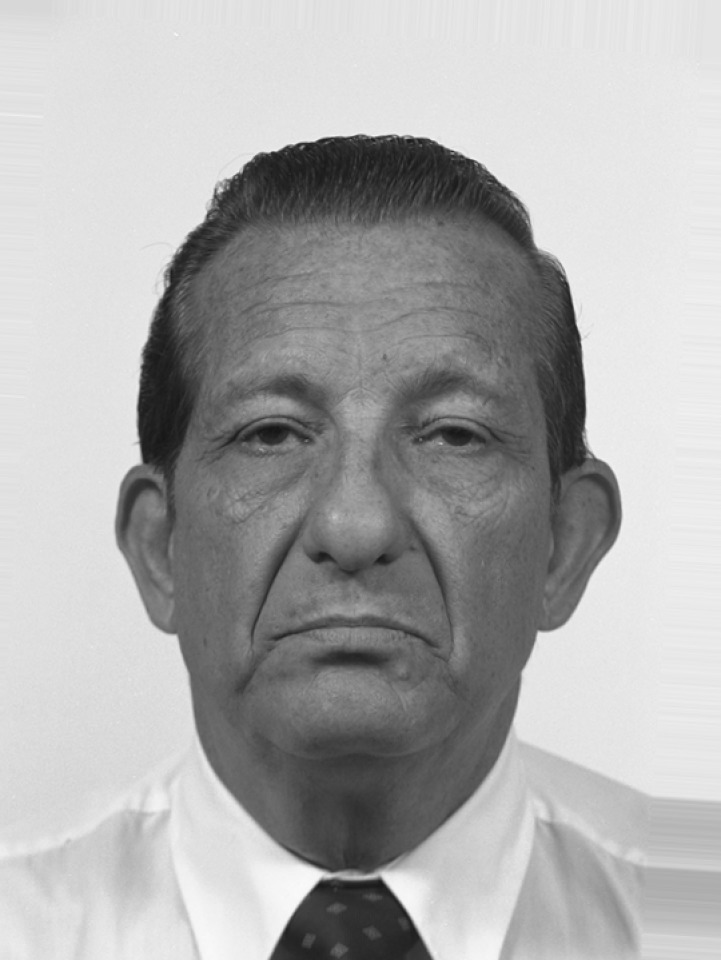}}	\hfill		
	\caption{Attack presentation classification error examples on FERET.}
	\label{fig:apce-feret}
	\vspace{-0.5cm}
\end{figure}

\begin{figure}
	\hfill
	\subfloat[Reference 1]{\includegraphics[height = 2.8cm]{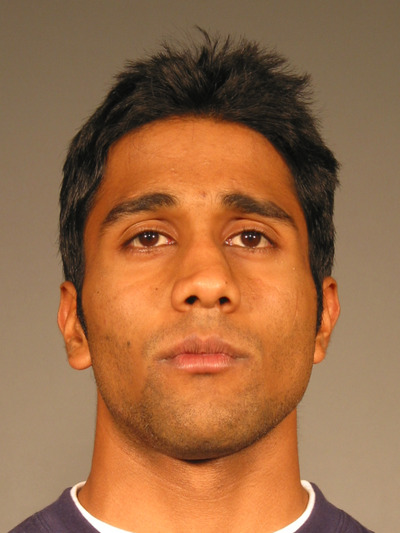}}\hfill
	\subfloat[Probe 1]{\includegraphics[height = 2.8cm]{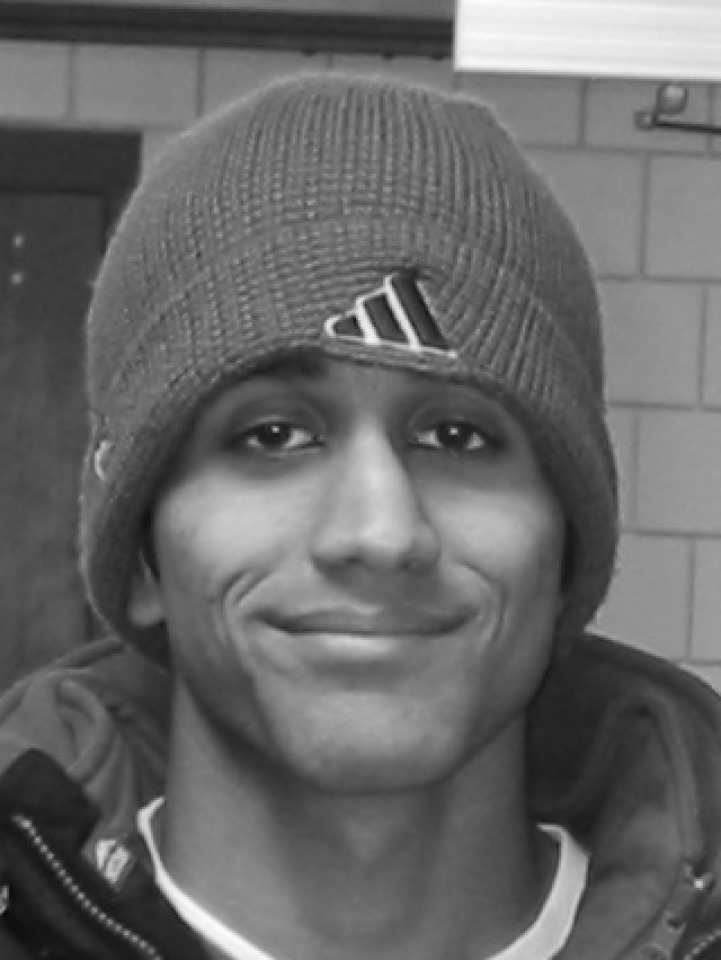}}\hfill
	\subfloat[Reference 2]{\includegraphics[height = 2.8cm]{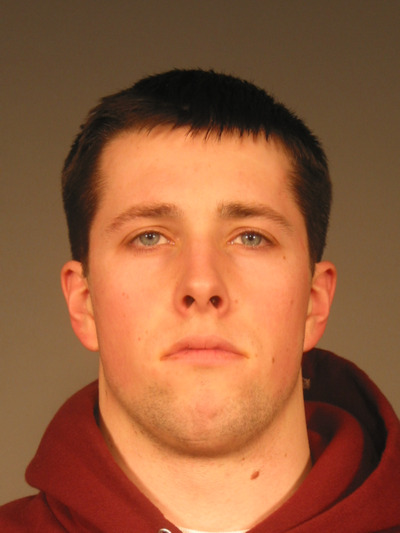}}\hfill
	\subfloat[Probe 2]{\includegraphics[height = 2.8cm]{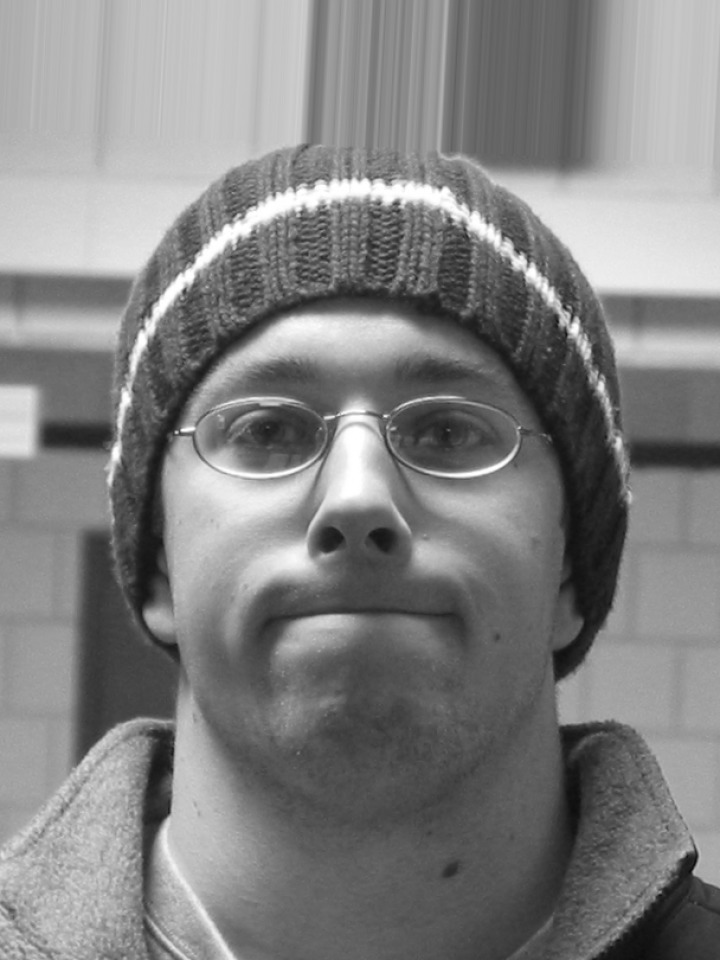}}	\hfill		
	\caption{Bona fide presentation classification error examples on FRGCv2.}
	\label{fig:bpcer-frgc}
	\vspace{-0.5cm}	
\end{figure}

It is noticeable that when evaluated on FRGCv2, some bona fide samples are very clearly, i.e., with rather high scores, misclassified as MAs, visible in Figure~\ref{fig:performance-pdf} (a) to (d) as a peak for bona fides at 1. On the contrary, on FERET, misclassifications occur mostly for MAs, which can be seen in Figure~\ref{fig:performance-pdf} (e) to (h) as small bumps in the MA score distributions close to 0. Examples for the above mentioned errors are given in Figure~\ref{fig:apce-feret} and Figure~\ref{fig:bpcer-frgc}. If the morph is of good quality and very similar to the sample, the MAD algorithm can no longer detect the MA. For example, in the samples shown in Figure~\ref{fig:apce-feret} it is difficult even for an experienced observer to detect the MA. Examples for wrongly classified bona fide authentication attempts are given in Figure~\ref{fig:bpcer-frgc}. The cause of this error is obvious. In these examples, the strong variance in facial expression and the presence of headgear in the probe images are particularly noticeable. However, such a variance can also be expected in a realistic border control scenario.

Due to the high certainty with which the algorithm assigns some samples to the wrong category, it is hardly possible to correct these errors by adjusting the MAD decision threshold value. Still, for both databases the score distributions of MAs and bona fide authentication attempts are very clearly separated. Thus, the error rates are quite stable within a considerable range of MAD decision threshold values (operating points), which makes the MAD algorithm very suitable for a practical application.

\subsection{NIST Face Recognition Vendor Test MORPH}
The NIST FRVT MORPH test provides ongoing independent testing of prototype face MAD methods.  The evaluation is designed to obtain commonly measured assessment of morph detection capability to inform developers and end-users. The test opened in June 2018, and, at the time of this writing, NIST has since received twelve different MAD algorithms. The latest report \cite{Ngan2019} includes morph detection performance results over thirteen datasets. On various unseen datasets where differential MAD was evaluated the proposed method achieved performance rates comparable to those presented in this work. Further, it outperformed all other methods by orders of magnitude, including implementations of single image MAD methods based on texture descriptors, forensic image analysis, and deep learning, i.e., \cite{Scherhag2018b,Ferrara2019,Ramachandra2017,Ramachandra2016,Debiasi2018,Debiasi2018a,Scherhag-PRNU-TBIOM-2019,Ramachandra19f}, as well as differential MAD methods employing texture descriptors and facial landmarks, i.e., \cite{Scherhag2018b,Damer2018b}. For detailed results the interested reader is referred to \cite{Ngan2019}. The submitted version of the proposed MAD method uses ArcFace for the extraction of deep face representations. For the training of the corresponding SVM, we utilized bona fide and morphed reference images from the FERET and FRGCv2 databases. In order to achieve high robustness morphed reference images generated by all morphing tools used in this work were considered in the training stage.    

\section{Conclusion}
\label{sec:conclusion}
Based on the experiments conducted in this work, the following conclusions are reached:
\begin{itemize}
	\item \textit{Detection performance}: the detection performance achieved by MAD based on deep face representations is promising and highly robust with respect to image post-processing, i.e., image compression, image resizing and even print-scan transformation. This is a clear advantage over MAD based on texture descriptors, which is typically quite sensitive to post-processing, particularly in more challenging scenarios. Moreover, the detection performance does not significantly depend on the post-processing applied to the training set, so that no scanned images are necessary for training.
	\item \textit{Heterogeneous morphing algorithms:} morphs generated by morphing algorithms which produce obvious artefacts, e.g., clearly visible ghost artefacts, are generally detected with higher accuracy. Furthermore, the recognition performance slightly degrades if training and evaluation sets contain morphs generated by different morphing algorithms.
	\item \textit{Heterogeneous databases:} if training and testing is conducted on heterogeneous face image databases which contain face images with different conditions, e.g., variations in pose and lightning, detection performance is negatively affected. On databases obtained from subsets of the publicly available FERET and the FRGCv2 face database, experiments revealed higher detection accuracy on the FERET subset in which probe images only contain slight variations in expression and pose as opposed to the FRGCv2 subset, which additionally comprises probe images with variations in lightning and focus. It can be concluded that strong variations in lightning and focus of probe images represent especially challenging conditions for differential MAD.
	\item \textit{Machine learning-based classifiers:} among the tested machine learning-based classifiers, i.e., AdaBoost, Gradient Boosting, Random Forest and Support Vector Machine (SVM), SVM-based classifiers generally revealed most competitive detection performance across the vast majority of conducted experiments.
	\item \textit{Commercial vs. open-source:} while commercial face recognition algorithms frequently outperform corresponding open-source implementations, this is not necessarily the case for MAD. Precisely, for the task of MAD, deep face representations obtained from open-source algorithms, e.g., ArcFace, might be better suited, compared to deep features extracted by commercial face recognition systems.	
\end{itemize}

Furthermore, this work underlines the need for realistic databases. Not only the quality of the reference, but also the quality of the probes has a strong influence on the detection performance of the MAD algorithms. Therefore, for the development of algorithms, which are deployable in a real world scenario, it is necessary to test on realistic data. However, there is still the problem that the exchange of databases is difficult due to privacy regulations.

\section*{Acknowledgment}
This research work has been funded by the German Federal Ministry  of  Education  and  Research  and  the  Hessen  StateMinistry for Higher Education, Research and the Arts within their joint support of the National Research Center for Applied  Cybersecurity  ATHENE  as well as by the Federal Office of Information Security (BSI) within the FACETRUST project. The UBO-Morpher as well as the Demorphing implementation was kindly provided by the University of Bologna.

\ifCLASSOPTIONcaptionsoff
  \newpage
\fi




\bibliographystyle{IEEEtran}
\bibliography{IEEEabrv,bibliography}

%
%
%

%

\begin{IEEEbiography}[{\includegraphics[width=1in,height=1.25in,clip,keepaspectratio]{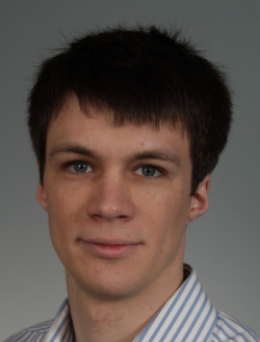}}]{Ulrich Scherhag}
	received his B.Eng degree (Electrical Engineering) in 2012 from the Duale Hochschule Baden-Württemberg, Mannheim. He stated studying computer science in 2014 at Hochschule Darmstadt and received the M.Sc degree (Computer Science, IT-Security) in 2016, for which he was granted the CAST Award IT-Security 2016. Since 2016 he is a Ph.D. Student Member of da/sec at the Center for Research in Security and Privacy (CRISP). He is a member of the European Association for Biometrics (EAB) and a Reviewer for the International Conference of the Biometrics Special Interest Group (BIOSIG) and IEEE Access. His current research focuses on presentation attack detection and morphed face detection.
\end{IEEEbiography}

\begin{IEEEbiography}[{\includegraphics[width=1in,height=1.25in,clip,keepaspectratio]{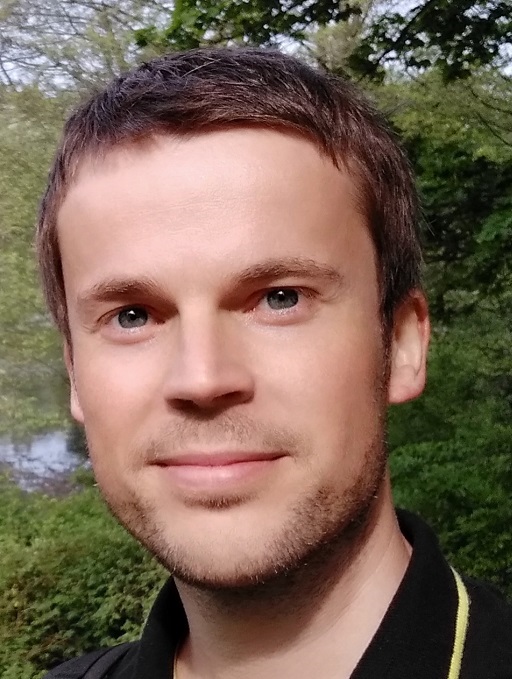}}]{Dr. Christian Rathgeb}
	is a Senior Researcher with the Faculty of Computer Science, Hochschule Darmstadt (HDA), Germany. He is a Principal Investigator in the Center for Research in Security and Privacy (CRISP). His research includes pattern recognition, iris and face recognition, security aspects of biometric systems, secure process design and privacy enhancing technologies for biometric systems. He co-authored over 100 technical papers in the field of biometrics. He is a winner of the EAB - European Biometrics Research Award 2012, the Austrian Award of Excellence 2012, Best Poster Paper Awards (IJCB'11, IJCB'14, ICB'15) and the Best Paper Award Bronze (ICB'18). He is a member of the European Association for Biometrics (EAB), a Program Chair of the International Conference of the Biometrics Special Interest Group (BIOSIG) and a editorial board member of IET Biometrics (IET BMT). He has served for various program committees and conferences (e.g. ICB, IJCB, BIOSIG, IWBF) and journals  as a reviewer (e.g. IEEE TIFS, IEEE TBIOM, IET BMT).
\end{IEEEbiography}

\begin{IEEEbiography}[{\includegraphics[width=1in,height=1.25in,clip,keepaspectratio]{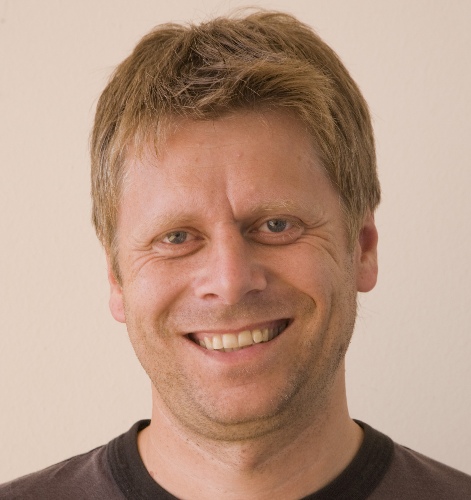}}]{Dr. Johannes Merkle}
	Johannes Merkle graduated 1995 in mathematics at the University in Frankfurt. After working there for 3 years in the field of cryptography, he received his PhD in mathematics in 2000. Since 1998, he is working for as a security consultant at secunet Security Networks, where he focussed on cryptography, public key infrastructures, and biometric cryptosystems. He has published several research papers on biometric template protection, fingerprint recognition and elliptic curve cryptography.
\end{IEEEbiography}

\begin{IEEEbiography}[{\includegraphics[width=1in,height=1.25in,clip,keepaspectratio]{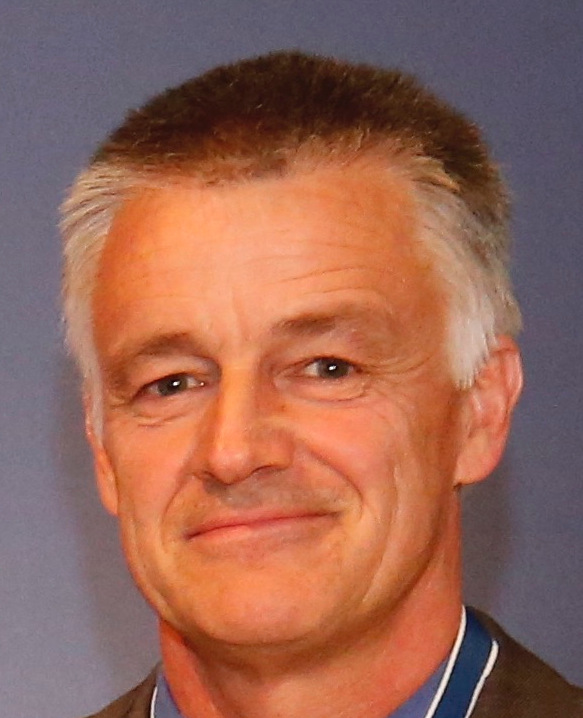}}]{Prof. Dr. Christoph Busch}
	is member of the Department of Information Security and Communication Technology (IIK) at the Norwegian University of Science and Technology (NTNU), Norway. He holds a joint appointment with the computer science faculty at Hochschule Darmstadt (HDA), Germany. Further he lectures Biometric Systems at Technical University of Denmark (DTU) since 2007.
	Christoph Busch co-authored more than 400 technical papers and has been a speaker at international conferences. He served for various program committees (NIST IBPC, ICB, ICHB, BSI-Congress, GI-Congress, DACH, WEDELMUSIC, EUROGRAPHICS) and served for several conferences, journals and magazines as reviewer (e.g. ACM-SIGGRAPH, ACM-TISSEC, IEEE CG\&A, IEEE Transactions on Signal Processing, on Information Forensics and Security, on Pattern Analysis and Machine Intelligence and the Elsevier Journal Computers \& Security). He is also an appointed member of the editorial board of the IET journal on Biometrics and of IEEE TIFS journal. Furthermore, he chairs the biometrics working group of the TeleTrusT association as well as the German standardization body on Biometrics (DIN-NIA37). He is convenor of WG3 in ISO/IEC JTC1 SC37 on Biometrics and active member of CEN TC 224 WG18.
\end{IEEEbiography}

%
%




\end{document}